\input harvmac

\input amssym
\input epsf


\newfam\frakfam
\font\teneufm=eufm10
\font\seveneufm=eufm7
\font\fiveeufm=eufm5
\textfont\frakfam=\teneufm
\scriptfont\frakfam=\seveneufm
\scriptscriptfont\frakfam=\fiveeufm


\def\bb{
\font\tenmsb=msbm10
\font\sevenmsb=msbm7
\font\fivemsb=msbm5
\textfont1=\tenmsb
\scriptfont1=\sevenmsb
\scriptscriptfont1=\fivemsb
}



\newfam\dsromfam
\font\tendsrom=dsrom10
\textfont\dsromfam=\tendsrom
\def\ds{\fam\dsromfam \tendsrom}


\newfam\mbffam
\font\tenmbf=cmmib10
\font\sevenmbf=cmmib7
\font\fivembf=cmmib5
\textfont\mbffam=\tenmbf
\scriptfont\mbffam=\sevenmbf
\scriptscriptfont\mbffam=\fivembf


\newfam\mbfcalfam
\font\tenmbfcal=cmbsy10
\font\sevenmbfcal=cmbsy7
\font\fivembfcal=cmbsy5
\textfont\mbfcalfam=\tenmbfcal
\scriptfont\mbfcalfam=\sevenmbfcal
\scriptscriptfont\mbfcalfam=\fivembfcal


\newfam\mscrfam
\font\tenmscr=rsfs10
\font\sevenmscr=rsfs7
\font\fivemscr=rsfs5
\textfont\mscrfam=\tenmscr
\scriptfont\mscrfam=\sevenmscr
\scriptscriptfont\mscrfam=\fivemscr
\def\scr{\fam\mscrfam \tenmscr}




\def\tilde{\widetilde}
\def\t{\tilde}
\def\hat{\widehat}

\def\bar{\overline}
\def\b{\bar}
\def\bsq#1{{{\b{#1}}^{\lower 2.5pt\hbox{$\scriptstyle 2$}}}}
\def\bexp#1#2{{{\b{#1}}^{\lower 2.5pt\hbox{$\scriptstyle #2$}}}}
\def\dotexp#1#2{{{#1}^{\lower 2.5pt\hbox{$\scriptstyle #2$}}}}


\def\rt2{\sqrt{2}}
\def\half {{1 \over 2}}
\def\Re{\mathop{\rm Re}}
\def\Im{\mathop{\rm Im}}
\def\d{\partial}

\def\mod{{\rm mod}}

\def\Tr{\mathop{\rm Tr}}

\def\sign{\mathop{\rm sgn}}


\font\tenbifull=cmmib10
\font\tenbimed=cmmib7
\font\tenbismall=cmmib5
\textfont9=\tenbifull \scriptfont9=\tenbimed
\scriptscriptfont9=\tenbismall

\mathchardef\bbGamma="7000
\mathchardef\bbDelta="7001
\mathchardef\bbPhi="7002
\mathchardef\bbAlpha="7003
\mathchardef\bbXi="7004
\mathchardef\bbPi="7005
\mathchardef\bbSigma="7006
\mathchardef\bbUpsilon="7007
\mathchardef\bbTheta="7008
\mathchardef\bbPsi="7009
\mathchardef\bbOmega="700A
\mathchardef\bbalpha="710B
\mathchardef\bbbeta="710C
\mathchardef\bbgamma="710D
\mathchardef\bbdelta="710E
\mathchardef\bbepsilon="710F
\mathchardef\bbzeta="7110
\mathchardef\bbeta="7111
\mathchardef\bbtheta="7112
\mathchardef\bbiota="7113
\mathchardef\bbkappa="7114
\mathchardef\bblambda="7115
\mathchardef\bbmu="7116
\mathchardef\bbnu="7117
\mathchardef\bbxi="7118
\mathchardef\bbpi="7119
\mathchardef\bbrho="711A
\mathchardef\bbsigma="711B
\mathchardef\bbtau="711C
\mathchardef\bbupsilon="711D
\mathchardef\bbphi="711E
\mathchardef\bbchi="711F
\mathchardef\bbpsi="7120
\mathchardef\bbomega="7121
\mathchardef\bbvarepsilon="7122
\mathchardef\bbvartheta="7123
\mathchardef\bbvarpi="7124
\mathchardef\bbvarrho="7125
\mathchardef\bbvarsigma="7126
\mathchardef\bbvarphi="7127




\def\thetasq{\theta^2}


\def\CA{{\cal A}}
\def\CB{{\cal B}}

\def\CH{{\cal H}}

\def\CJ{{\cal J}}

\def\CM{{\cal M}}
\def\CN{{\cal N}}
\def\CO{{\cal O}}

\def\CR{{\cal R}}
\def\CS{{\cal S}}

\def\CV{{\cal V}}
\def\CW{{\cal W}}


\def\1{{\ds 1}}
\def\R{\hbox{$\bb R$}}

\def\Z{\hbox{$\bb Z$}}


\def\ep{\varepsilon}

\noblackbox

\def\unit{\relax{\rm 1\kern-.26em I}}
\def\nada{\relax{\rm 0\kern-.30em l}}
\def\tilde{\widetilde}
\def\t{\tilde}

\def\mod{{\rm mod}}

\noblackbox
\def\IL{\relax{\rm I\kern-.18em L}}
\def\IH{\relax{\rm I\kern-.18em H}}
\def\IR{\relax{\rm I\kern-.18em R}}
\def\IC{\relax\hbox{$\inbar\kern-.3em{\rm C}$}}
\def\IZ{\relax\ifmmode\mathchoice
{\hbox{\cmss Z\kern-.4em Z}}{\hbox{\cmss Z\kern-.4em Z}} {\lower.9pt\hbox{\cmsss Z\kern-.4em Z}}
{\lower1.2pt\hbox{\cmsss Z\kern-.4em Z}}\else{\cmss Z\kern-.4em Z}\fi}
\def\CM {{\cal M}}

\def\CN {{\cal N}}
\def\CR {{\cal R}}

\def\CJ {{\cal J}}
\def\partialslash{\not{\hbox{\kern-2pt $\partial$}}}

\def\CV {{\cal V}}
\def\CO {{\cal O}}

\def\CH {{\cal H}}

\def\CB {{\cal B}}
\def\CW{{\cal W}}
\def\CS {{\cal S}}
\def\CA{{\cal A}}

\def\CM {{\cal M}}
\def\CN {{\cal N}}

\def\CO {{\cal O}}

\def\CV{{\cal V }}

\def\CS {{\cal S }}

\def\Tr{{\rm Tr}}

\font\manual=manfnt \def\dbend{\lower3.5pt\hbox{\manual\char127}}

\def\IZ{\relax\ifmmode\mathchoice
{\hbox{\cmss Z\kern-.4em Z}}{\hbox{\cmss Z\kern-.4em Z}} {\lower.9pt\hbox{\cmsss Z\kern-.4em Z}}
{\lower1.2pt\hbox{\cmsss Z\kern-.4em Z}}\else{\cmss Z\kern-.4em Z}\fi}
\def\half {{1\over 2}}

\def\bar{\overline}
\def\CS{{\cal S}}
\def\CH{{\cal H}}

\def\rt2{\sqrt{2}}
\def\irt2{{1\over\sqrt{2}}}

\def\t{\tilde}
\def\hat{\widehat}
\def\slashchar#1{\setbox0=\hbox{$#1$}           
   \dimen0=\wd0                                 
   \setbox1=\hbox{/} \dimen1=\wd1               
   \ifdim\dimen0>\dimen1                        
      \rlap{\hbox to \dimen0{\hfil/\hfil}}      
      #1                                        
   \else                                        
      \rlap{\hbox to \dimen1{\hfil$#1$\hfil}}   
      /                                         
   \fi}

\def\foursqr#1#2{{\vcenter{\vbox{
    \hrule height.#2pt
    \hbox{\vrule width.#2pt height#1pt \kern#1pt
    \vrule width.#2pt}
    \hrule height.#2pt
    \hrule height.#2pt
    \hbox{\vrule width.#2pt height#1pt \kern#1pt
    \vrule width.#2pt}
    \hrule height.#2pt
        \hrule height.#2pt
    \hbox{\vrule width.#2pt height#1pt \kern#1pt
    \vrule width.#2pt}
    \hrule height.#2pt
        \hrule height.#2pt
    \hbox{\vrule width.#2pt height#1pt \kern#1pt
    \vrule width.#2pt}
    \hrule height.#2pt}}}}
\def\psqr#1#2{{\vcenter{\vbox{\hrule height.#2pt
    \hbox{\vrule width.#2pt height#1pt \kern#1pt
    \vrule width.#2pt}
    \hrule height.#2pt \hrule height.#2pt
    \hbox{\vrule width.#2pt height#1pt \kern#1pt
    \vrule width.#2pt}
    \hrule height.#2pt}}}}
\def\sqr#1#2{{\vcenter{\vbox{\hrule height.#2pt
    \hbox{\vrule width.#2pt height#1pt \kern#1pt
    \vrule width.#2pt}
    \hrule height.#2pt}}}}
\def\square{\mathchoice\sqr65\sqr65\sqr{2.1}3\sqr{1.5}3}

\def\figin{\epsfcheck\figin}\def\figins{\epsfcheck\figins}
\def\epsfcheck{\ifx\epsfbox\UnDeFiNeD
\message{(NO epsf.tex, FIGURES WILL BE IGNORED)}
\gdef\figin##1{\vskip2in}\gdef\figins##1{\hskip.5in}
\else\message{(FIGURES WILL BE INCLUDED)}%
\gdef\figin##1{##1}\gdef\figins##1{##1}\fi}
\def\DefWarn#1{}
\def\figinsert{\goodbreak\midinsert}
\def\ifig#1#2#3{\DefWarn#1\xdef#1{fig.~\the\figno}
\writedef{#1\leftbracket fig.\noexpand~\the\figno}%
\figinsert\figin{\centerline{#3}}\medskip\centerline{\vbox{\baselineskip12pt \advance\hsize by
-1truein\noindent\footnotefont{\bf Fig.~\the\figno:\ } \it#2}}
\bigskip\endinsert\global\advance\figno by1}

\lref\FrishmanDQ{
  Y.~Frishman, A.~Schwimmer, T.~Banks and S.~Yankielowicz,
  ``The Axial Anomaly And The Bound State Spectrum In Confining Theories,''
  Nucl.\ Phys.\  B {\bf 177}, 157 (1981).
}

\lref\JafferisUN{
  D.~L.~Jafferis,
  ``The Exact Superconformal R-Symmetry Extremizes Z,''
  arXiv:1012.3210 [hep-th].
}

\lref\DimofteJU{
  T.~Dimofte, D.~Gaiotto and S.~Gukov,
  ``Gauge Theories Labelled by Three-Manifolds,''
[arXiv:1108.4389 [hep-th]].
}

\lref\ColemanZI{
  S.~R.~Coleman and B.~R.~Hill,
  ``No More Corrections to the Topological Mass Term in QED in Three-Dimensions,''
Phys.\ Lett.\ B {\bf 159}, 184 (1985).
}

\lref\DumitrescuIU{
  T.~T.~Dumitrescu and N.~Seiberg,
  ``Supercurrents and Brane Currents in Diverse Dimensions,''
  JHEP {\bf 1107}, 095 (2011)
  [arXiv:1106.0031 [hep-th]].
}

\lref\BarnesBM{
  E.~Barnes, E.~Gorbatov, K.~A.~Intriligator, M.~Sudano and J.~Wright,
  ``The exact superconformal R-symmetry minimizes $\tau_{RR}$,''
  Nucl.\ Phys.\  B {\bf 730}, 210 (2005)
  [arXiv:hep-th/0507137].
}

\lref\SeibergQD{
  N.~Seiberg,
  ``Modifying the Sum Over Topological Sectors and Constraints on Supergravity,''
JHEP {\bf 1007}, 070 (2010).
[arXiv:1005.0002 [hep-th]].
}

\lref\WittenYA{
  E.~Witten,
  ``SL(2,Z) action on three-dimensional conformal field theories with Abelian symmetry,''
In *Shifman, M. (ed.) et al.: From fields to strings, vol. 2* 1173-1200.
[hep-th/0307041].
}

\lref\MaldacenaSS{
  J.~M.~Maldacena, G.~W.~Moore and N.~Seiberg,
  ``D-brane charges in five-brane backgrounds,''
JHEP {\bf 0110}, 005 (2001).
[hep-th/0108152].
}

\lref\GaiottoAK{
  D.~Gaiotto and E.~Witten,
  ``S-Duality of Boundary Conditions In N=4 Super Yang-Mills Theory,''
[arXiv:0807.3720 [hep-th]].
}

\lref\ZamolodchikovGT{
  A.~B.~Zamolodchikov,
  ``Irreversibility of the Flux of the Renormalization Group in a 2D Field Theory,''
JETP Lett.\  {\bf 43}, 730-732 (1986).}

\lref\JafferisZI{
  D.~L.~Jafferis, I.~R.~Klebanov, S.~S.~Pufu, B.~R.~Safdi,
  ``Towards the F-Theorem: N=2 Field Theories on the Three-Sphere,''
JHEP {\bf 1106}, 102 (2011).
[arXiv:1103.1181 [hep-th]].
}

\lref\JafferisNS{
  D.~Jafferis and X.~Yin,
  ``A Duality Appetizer,''
[arXiv:1103.5700 [hep-th]].
}

\lref\AlvarezGaumeIG{
  L.~Alvarez-Gaume and E.~Witten,
  ``Gravitational Anomalies,''
Nucl.\ Phys.\ B\ {\bf 234}, 269  (1984)..
}

\lref\FestucciaWS{
  G.~Festuccia and N.~Seiberg,
  ``Rigid Supersymmetric Theories in Curved Superspace,''
JHEP {\bf 1106}, 114 (2011).
[arXiv:1105.0689 [hep-th]].
}

\lref\SohniusTP{
  M.~F.~Sohnius and P.~C.~West,
  ``An Alternative Minimal Off-Shell Version of N=1 Supergravity,''
Phys.\ Lett.\ B {\bf 105}, 353 (1981).
}

\lref\BeniniMF{
  F.~Benini, C.~Closset and S.~Cremonesi,
  ``Comments on 3d Seiberg-like dualities,''
  JHEP {\bf 1110}, 075 (2011)
  [arXiv:1108.5373 [hep-th]].
}

\lref\BanksZN{
  T.~Banks and N.~Seiberg,
  ``Symmetries and Strings in Field Theory and Gravity,''
Phys.\ Rev.\ D {\bf 83}, 084019 (2011).
[arXiv:1011.5120 [hep-th]].
}

\lref\SeibergVC{
  N.~Seiberg,
  ``Naturalness versus supersymmetric nonrenormalization theorems,''
Phys.\ Lett.\ B {\bf 318}, 469 (1993).
[hep-ph/9309335].
}

\lref\AharonyBX{
  O.~Aharony, A.~Hanany, K.~A.~Intriligator, N.~Seiberg and M.~J.~Strassler,
  ``Aspects of N = 2 supersymmetric gauge theories in three dimensions,''
  Nucl.\ Phys.\  B {\bf 499}, 67 (1997)
  [arXiv:hep-th/9703110].
}

\lref\BelovZE{
  D.~Belov and G.~W.~Moore,
  ``Classification of Abelian spin Chern-Simons theories,''
[hep-th/0505235].
}

\lref\KapustinHK{
  A.~Kapustin and N.~Saulina,
  ``Topological boundary conditions in abelian Chern-Simons theory,''
Nucl.\ Phys.\ B {\bf 845}, 393 (2011).
[arXiv:1008.0654 [hep-th]].
}

\lref\RedlichKN{
  A.~N.~Redlich,
  ``Gauge Non-Invariance and Parity Non-Conservation of Three-Dimensional
  Fermions,''
  Phys.\ Rev.\ Lett.\  {\bf 52}, 18 (1984).
}

\lref\RedlichDV{
  A.~N.~Redlich,
  ``Parity Violation And Gauge Noninvariance Of The Effective Gauge Field
  Action In Three-Dimensions,''
  Phys.\ Rev.\  D {\bf 29}, 2366 (1984).
}

\lref\WittenHF{
  E.~Witten,
  ``Quantum field theory and the Jones polynomial,''
  Commun.\ Math.\ Phys.\  {\bf 121}, 351 (1989).
}

\lref\KapustinHA{
  A.~Kapustin and M.~J.~Strassler,
  ``On mirror symmetry in three-dimensional Abelian gauge theories,''
JHEP {\bf 9904}, 021 (1999).
[hep-th/9902033].
}

\lref\KapustinXQ{
  A.~Kapustin, B.~Willett and I.~Yaakov,
  ``Nonperturbative Tests of Three-Dimensional Dualities,''
JHEP {\bf 1010}, 013 (2010).
[arXiv:1003.5694 [hep-th]].
}

\lref\KapustinMH{
  A.~Kapustin, B.~Willett and I.~Yaakov,
  ``Tests of Seiberg-like Duality in Three Dimensions,''
[arXiv:1012.4021 [hep-th]].
}

\lref\WillettGP{
  B.~Willett and I.~Yaakov,
  ``N=2 Dualities and Z Extremization in Three Dimensions,''
[arXiv:1104.0487 [hep-th]].
}

\lref\CappelliYC{
  A.~Cappelli, D.~Friedan and J.~I.~Latorre,
  ``C Theorem And Spectral Representation,''
  Nucl.\ Phys.\  B {\bf 352}, 616 (1991).
}

\lref\KutasovXB{
  D.~Kutasov,
  ``Geometry On The Space Of Conformal Field Theories And Contact Terms,''
Phys.\ Lett.\ B {\bf 220}, 153 (1989).
}

\lref\ColemanYG{
  S.~R.~Coleman and B.~Grossman,
  ``'t Hooft's Consistency Condition as a Consequence of Analyticity and Unitarity,''
Nucl.\ Phys.\ B {\bf 203}, 205 (1982).
}

\lref\ClossetVG{
  C.~Closset, T.~T.~Dumitrescu, G.~Festuccia, Z.~Komargodski and N.~Seiberg,
  ``Contact Terms, Unitarity, and F-Maximization in Three-Dimensional Superconformal Theories,''
[arXiv:1205.4142 [hep-th]].
}

\lref\AharonyGP{
  O.~Aharony,
  ``IR duality in d = 3 N=2 supersymmetric USp(2N(c)) and U(N(c)) gauge theories,''
Phys.\ Lett.\ B {\bf 404}, 71 (1997).
[hep-th/9703215].
}

\lref\GiveonZN{
  A.~Giveon and D.~Kutasov,
  ``Seiberg Duality in Chern-Simons Theory,''
Nucl.\ Phys.\ B {\bf 812}, 1 (2009).
[arXiv:0808.0360 [hep-th]].
}

\lref\deBoerKR{
  J.~de Boer, K.~Hori and Y.~Oz,
  ``Dynamics of N=2 supersymmetric gauge theories in three-dimensions,''
Nucl.\ Phys.\ B {\bf 500}, 163 (1997).
[hep-th/9703100].
}

\lref\vdB{
  F.~van de Bult,
  ``Hyperbolic Hypergeometric Functions,''\vskip1pt
  [http://www.its.caltech.edu/~vdbult/Thesis.pdf].
}

\lref\AtiyahJF{
  M.~F.~Atiyah, V.~K.~Patodi and I.~M.~Singer,
  ``Spectral asymmetry and Riemannian Geometry 1,''
Math.\ Proc.\ Cambridge Phil.\ Soc.\  {\bf 77}, 43 (1975).
}

\lref\AtiyahJG{
  M.~F.~Atiyah, V.~K.~Patodi and I.~M.~Singer,
  ``Spectral asymmetry and Riemannian geometry 2,''
Math.\ Proc.\ Cambridge Phil.\ Soc.\  {\bf 78}, 405 (1976).
}

\lref\AtiyahJH{
  M.~F.~Atiyah, V.~K.~Patodi and I.~M.~Singer,
  ``Spectral asymmetry and Riemannian Geometry 3,''
Math.\ Proc.\ Cambridge Phil.\ Soc.\  {\bf 79}, 71 (1980).
}

\lref\NaculichNC{
  S.~G.~Naculich and H.~J.~Schnitzer,
  ``Level-rank duality of the U(N) WZW model, Chern-Simons theory, and 2-D qYM theory,''
JHEP {\bf 0706}, 023 (2007).
[hep-th/0703089 [HEP-TH]].
}

\lref\RocekBK{
  M.~Rocek and P.~van Nieuwenhuizen,
  ``$N \geq  2$ Supersymmetric Chern-Simons Terms As $d = 3$ Extended Conformal Supergravity,''
Class.\ Quant.\ Grav.\  {\bf 3}, 43 (1986).
}

\lref\KuzenkoXG{
  S.~M.~Kuzenko, U.~Lindstrom and G.~Tartaglino-Mazzucchelli,
  ``Off-shell supergravity-matter couplings in three dimensions,''
JHEP {\bf 1103}, 120 (2011).
[arXiv:1101.4013 [hep-th]].
}

\lref\KuzenkoRD{
  S.~M.~Kuzenko and G.~Tartaglino-Mazzucchelli,
  ``Three-dimensional N=2 (AdS) supergravity and associated supercurrents,''
JHEP {\bf 1112}, 052 (2011).
[arXiv:1109.0496 [hep-th]].
}

\lref\WittenKT{
  E.~Witten,
  ``Three-Dimensional Gravity Revisited,''
[arXiv:0706.3359 [hep-th]].
}

\lref\AchucarroVZ{
  A.~Achucarro and P.~K.~Townsend,
  ``A Chern-Simons Action for Three-Dimensional anti-de-Sitter Supergravity Theories,''
Phys.\ Lett.\ B {\bf 180}, 89 (1986).
}

\lref\AchucarroGM{
  A.~Achucarro and P.~K.~Townsend,
  ``Extended Supergravitites in $d = 2+1$ as Chern-Simons Theories,''
Phys.\ Lett.\ B {\bf 229}, 383 (1989).
}



\rightline{PUPT-2417}
\rightline{WIS/10/12-JUNE-DPPA}
\Title{
} {\vbox{\centerline{Comments on Chern-Simons Contact Terms}
\vskip7pt
\centerline{in Three Dimensions}
}}

\centerline{Cyril Closset,$^1$ Thomas T. Dumitrescu,$^{2}$ Guido Festuccia,$^3$}
\centerline{Zohar Komargodski,$^{1,3}$ and Nathan Seiberg\hskip1pt $^{3}$}
\bigskip
\centerline{ $^{1}$ {\it Weizmann Institute of Science, Rehovot
76100, Israel}}
 \centerline{$^{2}$ {\it Department of Physics, Princeton University, Princeton, NJ 08544, USA}}
  \centerline{$^{3}${\it
Institute for Advanced Study, Princeton, NJ 08540, USA}}

\vskip40pt

\noindent
We study contact terms of conserved currents and the energy-momentum tensor in three-dimensional quantum field theory.  They are associated with Chern-Simons terms for background fields.  While the integer parts of these contact terms are ambiguous, their fractional parts are meaningful physical observables.  In~$\CN=2$ supersymmetric theories with a~$U(1)_R$ symmetry some of these observables lead to an anomaly.
Moreover, they can be computed exactly using localization, leading to new tests of dualities.

\Date{June 2012}

\newsec{Introduction}

In quantum field theory, correlation functions of local operators may contain~$\delta$-function singularities at coincident points. Such contributions are referred to as contact terms. Typically, they are not universal. They depend on how the operators and coupling constants of the theory are defined at short distances, i.e.\ they depend on the regularization scheme. This is intuitively obvious, since contact terms probe the theory at very short distances, near the UV cutoff~$\Lambda$. If~$\Lambda$ is large but finite, correlation functions have features at distances of order~$\Lambda^{-1}$. In the limit~$\Lambda \rightarrow \infty$ some of these features can collapse into~$\delta$-function contact terms.

In this paper, we will discuss contact terms in two-point functions of conserved currents in three-dimensional quantum field theory. As we will see, they do not suffer from the scheme dependence of conventional contact terms, and hence they lead to interesting observables.

It is convenient to promote all coupling constants to classical background fields and specify a combined Lagrangian for the dynamical fields and the classical backgrounds. As an example, consider a scalar operator~$\CO(x)$, which couples to a classical background field~$\lambda(x)$,
\eqn\olag{{\scr L} = {\scr L}_0 + \lambda(x) \CO(x) + c \lambda^2(x) + c' \lambda(x) \d^2 \lambda(x) + \cdots~.}
Here~${\scr L}_0$ only depends on the dynamical fields and~$c, c'$ are constants. The ellipsis denotes other allowed local terms in~$\lambda(x)$. If the theory has a gap, we can construct a well-defined effective action~$F[\lambda]$ for the background field~$\lambda(x)$,
\eqn\fdef{e^{-F[\lambda]} = \left<e^{- \int d^3 x \, {\scr L}}\right>~,}
which captures correlation functions of~$\CO(x)$. (Since we are working in Euclidean signature, $F[\lambda]$ is nothing but the free energy.) At separated points, the connected two-point function~$\langle \CO(x) \CO(y)\rangle$ arises from the term in~\olag\ that is linear in~$\lambda(x)$. Terms quadratic in~$\lambda(x)$ give rise to contact terms:~$c \delta^{(3)} (x-y) + c' \d^2 \delta^{(3)}(x-y) + \cdots~$.

A change in the short-distance physics corresponds to modifying the Lagrangian~\olag\ by local counterterms in the dynamical and the background fields. For instance, we can change the constants~$c, c'$ by modifying the theory near the UV cutoff, and hence the corresponding contact terms are scheme dependent. Equivalently, a scheme change corresponds to a field redefinition of the coupling~$\lambda(x)$. This does not affect correlation functions at separated points, but it shifts the contact terms~\KutasovXB. A related statement concerns redundant operators, i.e.\ operators that vanish by the equations of motion, which have vanishing correlation functions at separated points but may give rise to non-trivial contact terms.

Nevertheless, contact terms are meaningful in several circumstances. For example, this is the case for contact terms associated with irrelevant operators, such as the magnetic moment operator. Dimensionless contact terms are also meaningful whenever some physical principle, such as a symmetry, restricts the allowed local counterterms. A well-known example is the seagull term in scalar electrodynamics, which is fixed by gauge invariance. Another example is the trace anomaly of the energy-momentum tensor~$T_{\mu\nu}$ in two-dimensional conformal field theories. Conformal invariance implies that~$T_\mu^\mu$ is a redundant operator. However, imposing the conservation law~$\d^\mu T_{\mu\nu} = 0$ implies that~$T_\mu^\mu$ has non-trivial contact terms. These contact terms are determined by the correlation functions of~$T_{\mu\nu}$ at separated points, and hence they are unambiguous and meaningful. This is typical of local anomalies~\refs{\FrishmanDQ\ColemanYG-\AlvarezGaumeIG}.

If we couple~$T_{\mu\nu}$ to a background metric~$g_{\mu\nu}$, the requirement that~$T_{\mu\nu}$ be conserved corresponds to diffeomorphism invariance, which restricts the set of allowed counterterms. In two dimensions, the contact terms of~$T_\mu^\mu$ are summarized by the formula~$\langle T_\mu^\mu\rangle ={c\over 24 \pi} R$, where~$c$ is the Virasoro central charge and~$R$ is the scalar curvature of the background metric.\foot{In our conventions, a~$d$-dimensional sphere of radius~$r$ has scalar curvature~$R = - {d (d-1) \over r^2}$~.} This result cannot be changed by the addition of diffeomorphism-invariant local counterterms.

The contact terms discussed above are either completely arbitrary or completely meaningful. In this paper we will discuss a third kind of contact term. Its integer part is scheme dependent and can be changed by adding local counterterms. However, its fractional part is an intrinsic physical observable.

Consider a three-dimensional quantum field theory with a global~$U(1)$ symmetry and its associated current~$j_\mu$. We will assume that the symmetry group is compact, i.e.\ only integer charges are allowed. The two-point function of~$j_\mu$ can include a contact term,
\eqn\twopc{\langle j_\mu(x) j_\nu(0)\rangle = \cdots  +{i\kappa\over 2\pi}\, \ep_{\mu\nu\rho} \partial^\rho\delta^{(3)}(x)~.}
Here~$\kappa$ is a real constant. Note that this term is consistent with current conservation. We can couple~$j_\mu$ to a background gauge field~$a_\mu$. The contact term in~\twopc\ corresponds to a Chern-Simons term for~$a_\mu$ in the effective action~$F[a]$,
\eqn\counterta{F[a] =  \cdots + {i \kappa \over 4 \pi} \int d^3 x \,  \ep^{\mu\nu\rho} a_\mu \d_\nu a_\rho~.}
We might attempt to shift~$\kappa \rightarrow \kappa + \delta\kappa$ by adding a Chern-Simons counterterm to the UV Lagrangian,
\eqn\countert{\delta {\scr L} =  {i \delta\kappa \over 4 \pi} \, \ep^{\mu\nu\rho} a_\mu \d_\nu a_\rho~.}
However, this term is not gauge invariant, and hence it is not a standard local counterterm.

We will now argue that~\countert\ is only a valid counterterm for certain quantized values of~$\delta\kappa$. Since counterterms summarize local physics near the cutoff scale, they are insensitive to global issues. Their contribution to the partition function~\fdef\ must be a well-defined, smooth functional for arbitrary configurations of the background fields and on arbitrary curved three-manifolds~$\CM_3$. Since we are interested in theories with fermions, we require~$\CM_3$ to be a spin manifold. Therefore~\countert\ is an admissible counterterm if its integral is a well-defined, smooth functional up to integer multiples of~$2 \pi i$. This restricts~$\delta\kappa$ to be an integer.

Usually, the quantization of~$\delta\kappa$ is said to follow from gauge invariance, but this is slightly imprecise. If the~$U(1)$ bundle corresponding to~$a_\mu$ is topologically trivial, then~$a_\mu$ is a good one-form. Since~\countert\ shifts by a total derivative under small gauge transformations, its integral is well defined. This is no longer the case for non-trivial bundles. In order to make sense of the integral, we extend~$a_\mu$ to a connection on a suitable~$U(1)$ bundle over a spin four-manifold~$\CM_4$ with boundary~$\CM_3$, and we define
\eqn\intcs{{i \over 4 \pi} \int_{\CM_3}  d^3 x\, \ep^{\mu\nu\rho} a_\mu \d_\nu a_\rho = {i \over 16 \pi} \int_{\CM_4} d^4 x \, \ep^{\mu\nu\rho\lambda} F_{\mu\nu} F_{\rho\lambda}~,}
where~$F_{\mu\nu} = \d_\mu a_\nu - \d_\nu a_\mu$ is the field strength. The right-hand side is a well-defined, smooth functional of~$a_\mu$, but it depends on the choice of~$\CM_4$. The difference between two choices~$\CM_4$ and~$\CM_4'$ is given by the integral over the closed four-manifold~$X_4$, which is obtained by properly gluing~$\CM_4$ and~$\CM_4'$ along their common boundary~$\CM_3$. Since~$X_4$ is also spin, we have
\eqn\xint{{i \over 16 \pi} \int_{X_4} d^4 x \, \ep^{\mu\nu\rho\lambda} F_{\mu\nu} F_{\rho\lambda} = 2 \pi i n~, \qquad n \in \Z~.}
Thus, if~$\delta\kappa$ is an integer, the integral of~\countert\ is well defined up to integer multiples of~$2 \pi i$.\foot{In a purely bosonic theory we do not require~$\CM_3$ to be spin. In this case~$\delta \kappa$ must be an even integer.}

We conclude that a counterterm of the from~\countert\ can only shift the contact term~$\kappa$ in~\twopc\ by an integer. Therefore, the fractional part~$\kappa \, \mod \, 1$ does not depend on short-distance physics. It is scheme independent and gives rise to a new meaningful observable in three-dimensional field theories. This observable is discussed in section~2.

In section~2, we will also discuss the corresponding observable for the energy-momentum tensor~$T_{\mu\nu}$. It is related to a contact term in the two-point function of~$T_{\mu\nu}$,
\eqn\ttct{\langle T_{\mu\nu} (x) T_{\rho \sigma}(0)\rangle = \cdots -{i \kappa_g \over 192 \pi} \left(\left(\ep_{\mu\rho\lambda} \d^\lambda (\d_\nu \d_\sigma - \d^2 \delta_{\nu\sigma}) + (\mu \leftrightarrow \nu)\right) + (\rho \leftrightarrow \sigma) \right)\delta^{(3)}(x)~.}
This contact term is associated with the gravitational Chern-Simons term, which is properly defined by extending the metric~$g_{\mu\nu}$ to a four-manifold,
\eqn\lcsdef{{i \over 192 \pi} \int_{\CM_3} \sqrt g \, d^3 x \, \ep^{\mu\nu\rho} \, \Tr \Big(\omega_\mu \d_\nu \omega_\rho + {2 \over 3} \omega_\mu \omega_\nu\omega_\rho\Big) = {i \over 768\pi} \int_{\CM_4} \sqrt g \, d^3 x \, \ep^{\mu\nu\rho\sigma} R_{\mu\nu\kappa\lambda} {R_{\rho\sigma}}^{\kappa\lambda}~.}
Here~$\omega_\mu$ is the spin connection and~$R_{\mu\nu\rho\sigma}$ is the Riemann curvature tensor. Note that we do not interpret the left-hand side of~\lcsdef\ as a Chern-Simons term for the~$SO(3)$ frame bundle. (See for instance the discussion in~\WittenKT.) As above, two different extensions of~$\CM_3$ differ by the integral over a closed spin four-manifold~$X_4$,
\eqn\xrint{{i \over 768\pi} \int_{X_4} \sqrt g \, d^3 x \, \ep^{\mu\nu\rho\sigma} R_{\mu\nu\kappa\lambda} {R_{\rho\sigma}}^{\kappa\lambda} = 2 \pi i n~, \qquad n \in \Z~.}
Therefore, the gravitational Chern-Simons term~\lcsdef\ is a valid counterterm, as long as its coefficient is an integer.\foot{If~$\CM_3$ is not spin, then the coefficient of~\lcsdef\ should be an integer multiple of~$16$.} Consequently, the integer part of the contact term~$\kappa_{g}$ in~\ttct\ is scheme dependent, while the fractional part~$\kappa_g \, \mod \, 1$ gives rise to a meaningful observable.

We would briefly like to comment on another possible definition of Chern-Simons counterterms, which results in the same quantization conditions for their coefficients. It involves the Atiyah-Patodi-Singer~$\eta$-invariant~\refs{\AtiyahJF\AtiyahJG-\AtiyahJH}, which is defined in terms of the eigenvalues of a certain Dirac operator on~$\CM_3$ that couples to~$a_\mu$ and~$g_{\mu\nu}$. (Loosely speaking, it counts the number of eigenvalues, weighted by their sign.) Therefore,~$\eta[a,g]$ is intrinsically three-dimensional and gauge invariant. The Atiyah-Patodi-Singer theorem states that~$i \pi \eta[a, g]$ differs from the four-dimensional integrals in~\intcs\ and~\lcsdef\ by an integer multiple of~$2 \pi i$. Hence, its variation gives rise to contact terms of the form~\twopc\ and~\ttct. Although~$\eta[a,g]$ is well defined, it jumps discontinuously by~$2$ when an eigenvalue of its associated Dirac operator crosses zero. Since short-distance counterterms should not be sensitive to zero-modes, we only allow~$i \pi \eta[a,g]$ with an integer coefficient.

In section~3, we discuss the observables~$\kappa \, \mod \, 1$ and~$\kappa_g \, \mod \, 1$ in several examples. We use our understanding of these contact terms to give an intuitive proof of a non-renormalization theorem due to Coleman and Hill~\ColemanZI.

In section 4 we extend our discussion to three-dimensional theories with~$\CN=2$ supersymmetry. Here we must distinguish between~$U(1)$ flavor symmetries and~$U(1)_R$ symmetries. Some of the contact terms associated with the~$R$-current are not consistent with conformal invariance. As we will see in section~5, this leads to a new anomaly in~$\CN=2$ superconformal theories, which is similar to the framing anomaly of~\WittenHF. The anomaly can lead to violations of conformal invariance and unitarity when the theory is placed on curved manifolds.

In section~6, we explore these phenomena in~$\CN=2$ supersymmetric QED (SQED) with a dynamical Chern-Simons term. For some range of parameters, this model is accessible in perturbation theory.

In supersymmetric theories, the observables defined in section~4 can be computed exactly using localization~\ClossetVG. In section~7, we compute them in several theories that were conjectured to be dual, subjecting these dualities to a new test.

Appendix~A contains simple free-field examples. In appendix~B we summarize relevant aspects of~$\CN=2$ supergravity.

\newsec{Two-Point Functions of Conserved Currents in Three Dimensions}

In this section we will discuss two-point functions of flavor currents and the energy-momentum tensor in three-dimensional quantum field theory, and we will explain in detail how the contact terms in these correlators give rise to a meaningful observable.

\subsec{Flavor Currents}

We will consider a~$U(1)$ flavor current~$j_\mu$. The extension to multiple~$U(1)$'s or to non-Abelian symmetries is straightforward. Current conservation restricts the two-point function of~$j_\mu$. In momentum space,\foot{Given two operators~$\CA(x)$ and~$\CB(x)$, we define~$\langle \CA(p) \CB(-p)\rangle = \int d^3 x \, e^{i p \cdot x} \, \langle \CA(x) \CB(0)\rangle~.$}
\eqn\twopt{\langle j_\mu(p) j_\nu(-p)\rangle =\tau\left({p^2\over\mu^2}\right) { p_\mu p_\nu - p^2 \delta_{\mu\nu} \over 16|p|}\   +\kappa\left({p^2\over\mu^2}\right) {\ep_{\mu\nu\rho} p^\rho\over 2\pi}~.}
Here~$\tau\left({p^2/\mu^2}\right)$ and~$\kappa\left({p^2/\mu^2}\right)$ are real, dimensionless structure functions and~$\mu$ is an arbitrary mass scale.

In a conformal field theory (CFT),~$\tau = \tau_{\rm CFT}$ and~$\kappa = \kappa_{\rm CFT}$ are independent of~$p^2$. (We assume throughout that the symmetry is not spontaneously broken.) In this case~\twopt\ leads to the following formula in position space:\foot{A term proportional to~$\ep_{\mu\nu\rho} \partial^\rho |x|^{-{3}}$, which is conserved and does not vanish at separated points, is not consistent with conformal invariance.}
\eqn\twopotp{\langle j_\mu(x) j_\nu(0)\rangle = \left(  \delta_{\mu\nu}\partial^2 - \partial_\mu \partial_\nu \right) {\tau_{\rm CFT} \over  32 \pi^2 x^2}   +{i\kappa_{\rm CFT}\over 2\pi} \ep_{\mu\nu\rho} \partial^\rho\delta^{(3)}(x) ~.}
This makes it clear that~$\tau_{\rm CFT}$ controls the behavior at separated points, while the term proportional to~$\kappa_{\rm CFT}$ is a pure contact term of the form~\twopc.  Unitarity implies that~$\tau_{\rm CFT} \geq 0$. If~$\tau_{\rm CFT} = 0$ then~$j_\mu$ is a redundant operator.

If the theory is not conformal, then~$\kappa\left({p^2/\mu^2}\right)$ may be a non-trivial function of~$p^2$. In this case the second term in~\twopt\ contributes to the two-point function at separated points, and hence it is manifestly physical. Shifting~$\kappa\left({p^2/\mu^2}\right)$ by a constant~$\delta\kappa$ only affects the contact term~\twopc. It corresponds to shifting the Lagrangian by the Chern-Simons counterterm~\countert. As explained in the introduction, shifts with arbitrary~$\delta\kappa$ may not always be allowed. We will return to this issue below.

It is natural to define the UV and IR values
\eqn\kappaUVIR{\kappa_{\rm  UV} = \lim_{p^2\to \infty} \kappa\left({p^2\over\mu^2}\right)~, \qquad \kappa_{\rm IR} = \lim_{p^2\to 0} \kappa\left({p^2\over\mu^2}\right) ~.}
Adding the counterterm~\countert\ shifts~$\kappa_{\rm UV}$ and~$\kappa_{\rm IR}$ by~$\delta\kappa$. Therefore~$\kappa_{\rm UV}-\kappa_{\rm IR}$ is not modified, and hence it is a physical observable.

We will now assume that the~$U(1)$ symmetry is compact, i.e.\ only integer charges are allowed. (This is always the case for theories with a Lagrangian description, as long as we pick a suitable basis for the Abelian flavor symmetries.) In this case, the coefficient~$\delta\kappa$ of the Chern-Simons counterterm~\countert\ must be an integer. Therefore, the entire fractional part~$\kappa(p^2/\mu^2) \, \mod \, 1$ is scheme independent. It is a physical observable for every value of~$p^2$. In particular, the constant~$\kappa_{\rm CFT} \, \mod \, 1$ is an intrinsic physical observable in any CFT.

The fractional part of~$\kappa_{\rm CFT}$ has a natural bulk interpretation for CFTs with an~$AdS_4$ dual. While the constant~$\tau_{\rm CFT}$ is related to the coupling of the bulk gauge field corresponding to~$j_\mu$, the fractional part of~$\kappa_{\rm CFT}$ is related to the bulk~$\theta$-angle. The freedom to shift~$\kappa_{\rm CFT}$ by an integer reflects the periodicity of~$\theta$, see for instance~\WittenYA.

In order to calculate the observable~$\kappa_{\rm CFT} \, \mod\, 1$ for a given CFT, we can embed the CFT into an RG flow from a theory whose~$\kappa$ is known -- for instance a free theory. We can then unambiguously calculate~$\kappa(p^2/\mu^2)$ to find the value of~$\kappa_{\rm CFT}$ in the IR. This procedure is carried out for free massive theories in appendix~A. More generally, if the RG flow is short, we can calculate the change in~$\kappa$ using (conformal) perturbation theory. In certain supersymmetric theories it is possible to calculate~$\kappa_{\rm CFT} \, \mod \, 1$ exactly using localization~\ClossetVG. This will be discussed in section~7.

We would like to offer another perspective on the observable related to~$\kappa(p^2)$. Using~\twopt, we can write the difference~$\kappa_{\rm UV}-\kappa_{\rm IR}$ as follows:
\eqn\diffi{\kappa_{\rm UV}-\kappa_{\rm IR} = {i \pi \over 6} \int_{\R^3-\{0\}} \, d^3 x \, x^2 \, \ep^{\mu\nu\rho}\, \d_\mu \langle j_\nu(x) j_\rho(0)\rangle~.}
The integral over~$\R^3-\{0\}$ excludes a small ball around~$x=0$, and hence it is not sensitive to contact terms.  The integral converges because the two-point function~$\ep^{\mu\nu\rho} \d_\mu \langle j_\nu(x) j_\rho(0)\rangle$ vanishes at separated points in a conformal field theory, so that it decays faster than~$1\over x^3$ in the IR and diverges more slowly than~$1\over x^3$ in the UV. Alternatively, we can use Cauchy's theorem to obtain the dispersion relation
\eqn\disp{\kappa_{\rm UV}-\kappa_{\rm IR} ={1\over \pi} \int_0^\infty {ds\over s} \, \Im\kappa\left(-{s\over \mu^2}\right)~.}
This integral converges for the same reasons as~\diffi. Since it only depends on the imaginary part of~$\kappa(p^2/\mu^2)$, it is physical.

The formulas~\diffi\ and~\disp\ show that the difference between~$\kappa_{\rm UV}$ and~$\kappa_{\rm IR}$ can be understood by integrating out massive degrees of freedom as we flow from the UV theory to the IR theory. Nevertheless, they capture the difference between two quantities that are intrinsic to these theories. Although there are generally many different RG flows that connect a pair of UV and IR theories, the integrals in~\diffi\ and~\disp\ are invariant under continuous deformations of the flow. This is very similar to well-known statements about the Virasoro central charge~$c$ in two dimensions. In particular, the sum rules~\diffi\ and~\disp\ are analogous to the sum rules in~\refs{\ZamolodchikovGT,\CappelliYC} for the change in~$c$ along an RG flow.

\subsec{Energy-Momentum Tensor}

We can repeat the analysis of the previous subsection for the two-point function of the energy-momentum tensor~$T_{\mu\nu}$, which depends on three dimensionless structure functions~$\tau_g(p^2/\mu^2)$, $\tau_g'(p^2/\mu^2)$, and~$\kappa_g(p^2/\mu^2)$,
\eqn\emtensor{\eqalign{\langle T_{\mu\nu} (p) T_{\rho\sigma}(-p)\rangle &= -(p_\mu p_\nu - p^2 \delta_{\mu\nu})(p_\rho p_\sigma - p^2 \delta_{\rho\sigma}) {\tau_g\left({p^2 / \mu^2}\right)  \over |p|}  \cr
& - \left((p_\mu p_\rho - p^2 \delta_{\mu\rho})(p_\nu p_\sigma - p^2 \delta_{\nu\sigma}) + (\mu \leftrightarrow \nu)\right) {\tau_g'\left({p^2 / \mu^2}\right) \over |p|} \cr
& + { \kappa_g\left({p^2 / \mu^2}\right) \over 192 \pi}\left(\left(\ep_{\mu\rho\lambda} p^\lambda (p_\nu p_\sigma - p^2 \delta_{\nu\sigma}) + (\mu \leftrightarrow \nu) \right) + (\rho \leftrightarrow \sigma)\right)~.}}
Unitarity implies that~$\tau_g(p^2/\mu^2) + \tau_g'(p^2/\mu^2) \geq 0$. If the equality is saturated, the trace~$T_\mu^\mu$ becomes a redundant operator. This is the case in a CFT, where~$\tau_g = -\tau'_g$ and~$\kappa_g$ are constants. The terms proportional to~$\tau_g$ determine the correlation function at separated points. The term proportional to~$\kappa_g$ gives rise to a conformally invariant contact term~\ttct. It is associated with the gravitational Chern-Simons term~\lcsdef, which is invariant under a conformal rescaling of the metric. Unlike the Abelian case discussed above, the contact term~$\kappa_g$ is also present in higher-point functions of~$T_{\mu\nu}$. (This is also true for non-Abelian flavor currents.)

Repeating the logic of the previous subsection, we conclude that~$\kappa_{g, {\rm UV}} - \kappa_{g, {\rm IR}}$ is physical and can in principle be computed along any RG flow. Moreover, the quantization condition on the coefficient of the gravitational Chern-Simons term~\lcsdef\ implies that the fractional part~$\kappa_g(p^2/\mu^2) \, \mod \, 1$ is a physical observable for any value of~$p^2$. In particular~$\kappa_{g, {\rm CFT}} \, \mod \, 1$ is an intrinsic observable in any CFT.

\newsec{Examples}

In this section we discuss a number of examples that illustrate our general discussion above.  An important example with~$\CN=2$ supersymmetry will be discussed in section~6.  Other examples with~$\CN=4$ supersymmetry appear in~\GaiottoAK.

\subsec{Free Fermions}

We begin by considering a theory of~$N$ free Dirac fermions of charge~$+1$ with real masses~$m_i$. Here we make contact with the parity anomaly of~\refs{\RedlichKN,\RedlichDV,\AlvarezGaumeIG}. As is reviewed in appendix~A, integrating out a Dirac fermion of mass~$m$ and charge~$+1$ shifts~$\kappa$ by~$-\half \sign (m)$, and hence we find that
\eqn\freef{\kappa_{\rm UV} - \kappa_{\rm IR} = \half\sum_{i=1}^N \sign \left(m_i\right)~.}
If~$N$ is odd, this difference is a half-integer. Setting~$\kappa_{\rm UV} = 0$ implies that~$\kappa_{\rm IR}$ is a half-integer, even though the IR theory is empty. In the introduction, we argued that short-distance physics can only shift~$\kappa$ by an integer. The same argument implies that~$\kappa_{\rm IR}$ must be an integer if the IR theory is fully gapped.\foot{We refer to a theory as fully gapped when it does not contain any massless or topological degrees of freedom.} We conclude that it is inconsistent to set~$\kappa_{\rm UV}$ to zero; it must be a half-integer. Therefore,
\eqn\kappaFF{\eqalign{&\kappa_{\rm UV}=\half + n~, \qquad n\in \Z~,\cr
&\kappa_{\rm IR}= \kappa_{\rm UV} - \half\sum_{i=1}^N \sign (m_i) \in \Z~.}}
The half-integer value of~$\kappa_{\rm UV}$ implies that the UV theory is not parity invariant, even though it does not contain any parity-violating mass terms. This is known as the parity anomaly~\refs{\RedlichKN,\RedlichDV,\AlvarezGaumeIG}.

We can use~\kappaFF\ to find the observable~$\kappa_{\rm CFT} \, \mod \, 1$ for the CFT that consists of~$N$ free massless Dirac fermions of unit charge:
\eqn\obsdir{\kappa_{\rm CFT} \, \mod \, 1 = \cases{0 \qquad N \;  {\rm even} \cr
\half \qquad N \; {\rm odd}}}
This illustrates the fact that we can calculate~$\kappa_{\rm CFT}$, if we can connect the CFT of interest to a theory with a known value of~$\kappa$. Here we used the fact that the fully gapped IR theory has integer~$\kappa_{\rm IR}$.

We can repeat the above discussion for the contact term~$\kappa_g$ that appears in the two-point function of the energy-momentum tensor. Integrating out a Dirac fermion of mass~$m$ shifts~$\kappa_g$ by~$-\sign (m)$, so that
\eqn\difkl{\kappa_{g, {\rm UV}}-\kappa_{g, {\rm IR}}= \sum_i \sign (m_i)~.}
If we instead consider~$N$ Majorana fermions with masses~$m_i$, then~$\kappa_{g, {\rm UV}} - \kappa_{g, {\rm IR}}$ would be half the answer in~\difkl. Since~$\kappa_{g, {\rm IR}}$ must be an integer in a fully gapped theory, we conclude that~$\kappa_{g, {\rm UV}}$ is a half-integer if the UV theory consists of an odd number of massless Majorana fermions.  This is the gravitational analogue of the parity anomaly.

\subsec{Topological Currents and Fractional Values of~$\kappa$}

Consider a dynamical~$U(1)$ gauge field~$A_\mu$, and the associated topological current
\eqn\topcon{j_\mu={i p\over 2\pi} \, \ep_{\mu\nu\rho}  \d^\nu A^\rho~, \qquad p \in \Z~.}
Note that the corresponding charges are integer multiples of~$p$. We study the free topological theory consisting of two~$U(1)$ gauge fields -- the dynamical gauge field~$A_\mu$ and a classical background gauge field~$a_\mu$ -- with Lagrangian~\refs{\WittenYA,\MaldacenaSS\BelovZE\GaiottoAK\KapustinHK-\BanksZN}
\eqn\anaLag{{\scr L} = {i \over 4\pi} \left(k\, \ep^{\mu\nu\rho} A_\mu \d_\nu A_\rho + 2\, p\, \ep^{\mu\nu\rho} a_\mu \d_\nu A_\rho+ q \, \ep^{\mu\nu\rho} a_\mu  \d_\nu a_\rho \right)~,\qquad k, p, q \in \Z ~.}
The background field~$a_\mu$ couples to the topological current~$j_\mu$ in~\topcon.  In order to compute the contact term~$\kappa$ corresponding to~$j_\mu$, we naively integrate out the dynamical field~$A_\mu$ to obtain an effective Lagrangian for~$a_\mu$,
\eqn\alae{{\scr L}_{\rm eff}= {i \kappa \over 4\pi} \ep^{\mu\nu\rho} a_\mu \d_\nu a_\rho~, \qquad \kappa = q - {p^2\over  k}~.}

Let us examine the derivation of \alae\ more carefully.  The equation of motion for~$A_\mu$ is
\eqn\Aeom{k\ep^{\mu\nu\rho} \partial_\nu A_\rho = -p \ep^{\mu\nu\rho} \partial_\nu a_\rho~~.}
Assuming, for simplicity, that~$k$ and~$p$ are relatively prime, this equation can be solved only if the flux of~$a_\mu$ through every two-cycle is an integer multiple of~$k$. When this is not the case the functional integral vanishes.  If the fluxes of $a_\mu$ are multiples of~$k$, the derivation of \alae\ is valid.  For these configurations the fractional value of~$\kappa$ is harmless.

This example shows that~$\kappa$ is not necessarily an integer, even if the theory contains only topological degrees of freedom. Equivalently, the observable~$\kappa \, \mod \, 1$ is sensitive to topological degrees of freedom. We would like to make a few additional comments:
\item{1.)} The freedom in shifting the Lagrangian by a Chern-Simons counterterm~\countert\ with integer~$\delta\kappa$ amounts to changing the integer~$q$ in~\anaLag.
\item{2.)} The value~$\kappa=q - {p^2\over  k}$ can be measured by making the background field~$a_\mu$ dynamical and studying correlation functions of Wilson loops for~$a_\mu$ in flat Euclidean space~$\R^3$. These correlation functions can be determined using either the original theory~\anaLag\ or the effective Lagrangian~\alae.
\item{3.)} Consider a CFT that consists of two decoupled sectors: a nontrivial~CFT$_0$ with a global~$U(1)$ current~$j^{(0)}_\mu$ and a~$U(1)$ Chern-Simons theory with level~$k$ and topological current~${i p \over 2 \pi} \ep_{\mu\nu\rho} \d^\nu A^\rho$. We will study the linear combination~$j_\mu = j^{(0)}_\mu + {i p \over 2 \pi} \ep_{\mu\nu\rho} \d^\nu A^\rho$. Denoting the contact term in the two-point function of~$j_\mu^{(0)}$ by~$\kappa_0$, the contact term~$\kappa$ corresponding to~$j_\mu$ is given by
\eqn\kappamixed{\kappa=\kappa_0 - {p^2\over  k}+ ({\rm integer})~.}
Since the topological current is a redundant operator, it is not possible to extract~$\kappa$ by studying correlation functions of local operators at separated points. Nevertheless, the fractional part of~$\kappa$ is an intrinsic physical observable. This is an example of a general point that was recently emphasized in~\SeibergQD: a quantum field theory is not uniquely characterized by its local operators and their correlation functions at separated points. The presence of topological degrees of freedom makes it necessary to also study various extended objects, such as line or surface operators.

\subsec{A Non-Renormalization Theorem}

Consider an RG flow from a free theory in the UV to a fully gapped theory in the IR.
(Recall that a theory is fully gapped when it does not contain massless or topological degrees of freedom.)
In this case, we can identify~$\kappa_{\rm IR}$ with the coefficient of the Chern-Simons term for the background field~$a_\mu$ in the Wilsonian effective action.
Since the IR theory is fully gapped, $\kappa_{\rm IR}$ must be an integer. Depending on the number of fermions in the free UV theory, $\kappa_{\rm UV}$ is either an integer or a half-integer. Therefore, the difference~$\kappa_{\rm UV} - \kappa_{\rm IR}$ is either an integer or a half-integer, and hence it cannot change under smooth deformations of the coupling constants. It follows that this difference is only generated at one-loop. This is closely related to a non-renormalization theorem due to Coleman and Hill~\ColemanZI, which was proved through a detailed analysis of Feynman diagrams. Note that our argument applies to Abelian and non-Abelian flavor currents, as well as the  energy-momentum tensor.

When the IR theory has a gap, but contains some topological degrees of freedom, $\kappa$ need not be captured by the Wilsonian effective action. As in the previous subsection, it can receive contributions from the topological sector. If the flow is perturbative, we can distinguish 1PI diagrams. The results of~\ColemanZI\ imply that 1PI diagrams only contribute to~$\kappa$ associated with a flavor current at one-loop. (The fractional contribution discussed in the previous subsection arises from diagrams that are not 1PI.) However, this is no longer true for~$\kappa_g$, which is associated with the energy-momentum tensor. For instance, $\kappa_g$ receives higher loop contributions from 1PI diagrams in pure non-Abelian Chern-Simons theory~\WittenHF.

\subsec{Flowing Close to a Fixed Point}

Consider an RG flow with two crossover scales~$M\gg m$.  The UV consists of a free theory that is deformed by a relevant operator. Below the scale~$M$, the theory flows very close to a CFT. This CFT is further deformed by a relevant operator, so that it flows to a gapped theory below a scale~$m\ll M$.

If the theory has a~$U(1)$ flavor current~$j_\mu$, the structure functions in~\twopt\ interpolate between their values in the UV, through the CFT values, down to the IR:
\eqn\kappataup{\eqalign{
&\tau\approx \cases{
\tau_{\rm UV} & $p^2\gg M^2$ \cr
\tau_{\rm CFT} & $m^2\ll p^2 \ll M^2$\cr
\tau_{\rm IR}& $p^2\ll m^2$}
\cr
&\kappa \approx \cases{
\kappa_{\rm UV} & $p^2\gg M^2$ \cr
\kappa_{\rm CFT}  & $m^2\ll p^2 \ll M^2$\cr
\kappa_{\rm IR} & $p^2\ll m^2$}}}

Since the UV theory is free, $\tau_{\rm UV}$ is easily computed (see appendix~A). In a free theory we can always take the global symmetry group to be compact.  This implies that~$\kappa_{\rm UV}$ is either integer or half-integer, depending on the number of fermions that are charged under~$j_\mu$. If~$j_\mu$ does not mix with a topological current in the IR, then~$\tau_{\rm IR}$ vanishes and~$\kappa_{\rm IR}$ must be an integer. This follows from the fact that the theory is gapped.

Since we know~$\kappa_{\rm UV}$ and~$\kappa_{\rm IR}$, we can use the flow to give two complementary arguments that~$\kappa_{\rm CFT} \, \mod \, 1$ is an intrinsic observable of the CFT:
\item{1.)} {\it The flow from the UV to the CFT:}  Here we start with a well- defined~$\kappa_{\rm UV}$, which can only be shifted by an integer.  Since~$\kappa_{\rm UV}-\kappa_{\rm CFT}$ is physical, it follows that~$\kappa_{\rm CFT}$ is well defined modulo an integer.
\item{2.)} {\it The flow from the CFT to the IR:}  We can discuss the CFT without flowing into it from a free UV theory.  If the CFT can be deformed by a relevant operator such that it flows to a fully gapped theory, then~$\kappa_{\rm IR}$ must be an integer.  Since~$\kappa_{\rm CFT}-\kappa_{\rm IR}$ is physical and only depends on information intrinsic to the CFT, i.e.\ the relevant deformation that we used to flow out, we conclude that the fractional part of~$\kappa_{\rm CFT}$ is an intrinsic observable of the CFT.

\noindent Below, we will see examples of such flows, and we will use them to compute~$\kappa_{\rm CFT} \, \mod \, 1$.  For the theory discussed in section~6, we will check explicitly that flowing into or out of the CFT gives the same answer for this observable.

\newsec{Theories with~$\CN=2$ Supersymmetry}

In this section we extend the previous discussion to three-dimensional theories with~$\CN=2$ supersymmetry. Here we must distinguish between~$U(1)$ flavor symmetries and~$U(1)_R$ symmetries.

\subsec{Flavor Symmetries}

A~$U(1)$ flavor current~$j_\mu$ is embedded in a real linear superfield~$\CJ$, which satisfies~$D^2 \CJ = \b D^2 \CJ = 0$. In components,
\eqn\jcomp{\CJ = J + i \theta j + i \b \theta \b j + i \theta \b \theta K - \left(\theta \gamma^\mu \b \theta\right) j_\mu  - \half \thetasq \b \theta \gamma^\mu \d_\mu j - \half \b \thetasq \theta \gamma^\mu \d_\mu \b j + {1 \over 4} \thetasq \b \thetasq \d^2 J~.}
The supersymmetry Ward identities imply the following extension of~\twopt:\foot{Supersymmetry also fixes the two-point function of the fermionic operators~$j_\alpha$ and~$\b j_\alpha$ in terms of~$\hat \tau_{ff}$ and~$\kappa_{ff}$, but in order to simplify the presentation, we will restrict our discussion to bosonic operators.}
\eqn\threefun{\eqalign{& \langle j_\mu(p)j_\nu(-p)\rangle = (p_\mu p_\nu - p^2 \delta_{\mu\nu}) {\hat \tau_{ff} \over 8 |p|} + \ep_{\mu\nu\rho} p^\rho {\kappa_{ff}\over 2 \pi}~,\cr
& \langle J(p) J(-p) \rangle = {\hat \tau_{ff} \over 8 |p|}~,\cr
& \langle K(p) K(-p)\rangle  = - {|p|  \over 8} \, \hat \tau_{ff}~,\cr
& \langle J(p) K(-p) \rangle = {\kappa_{ff} \over 2 \pi}~.}}
Here we have defined~$\hat \tau_{ff} = \half \tau$, so that~$\hat \tau_{ff} =1$ for a free massless chiral superfield of charge~$+1$, and we have also renamed~$\kappa_{ff} = \kappa$. The subscript~$ff$ emphasizes the fact that we are discussing two-point functions of flavor currents.

As in the non-supersymmetric case, we can couple the flavor current to a background gauge field. Following~\refs{\SeibergVC,\AharonyBX}, we should couple~$\CJ$ to a background vector superfield,
\eqn\gaugemult{\CV = \cdots + \left(\theta \gamma^\mu \b \theta\right) a_\mu - i \theta \b \theta \sigma - i \thetasq \b \theta \b \lambda + i \b \thetasq \theta \lambda - \half \thetasq \b \thetasq D~.}
Background gauge transformations shift~$\CV \rightarrow \CV + \Lambda + \b \Lambda$ with chiral~$\Lambda$, so that~$\sigma$ and~$D$ are gauge invariant, while~$a_\mu$ transforms like an ordinary gauge field. (The ellipsis denotes fields that are pure gauge modes and do not appear in gauge-invariant functionals of~$\CV$.) The coupling of~$\CJ$ to~$\CV$ takes the form
\eqn\flatjv{\delta {\scr L} = - 2 \int d^4 \theta \, \CJ \CV =  - j_\mu a^\mu -  K\sigma - J D + ({\rm fermions})~.}
As before, it may be necessary to also add higher-order terms in~$\CV$ to maintain gauge invariance.

We can now adapt our previous discussion to~$\kappa_{ff}$. According to~\threefun, a constant value of~$\kappa_{ff}$ gives rise to contact terms in both~$\langle j_\mu(p)j_\nu(-p)\rangle$ and~$\langle J(p) K(-p) \rangle$. These contact terms correspond to a supersymmetric Chern-Simons term for the background field~$\CV$,
\eqn\ffcs{{\scr L}_{ff} = -{\kappa_{ff} \over 2 \pi}  \int d^4 \theta \, \Sigma\, \CV = {\kappa_{ff} \over 4 \pi} \left(i \ep^{\mu\nu\rho} a_\mu \d_\nu a_\rho - 2 \sigma D + \left({\rm fermions}\right)\right)~.}
Here the real linear superfield~$\Sigma = {i \over 2} \b D D \CV$ is the gauge-invariant field strength corresponding to~$\CV$. If the~$U(1)$ flavor symmetry is compact, then the same arguments as above imply that short-distance counterterms can only shift~$\kappa_{ff}$ by an integer, and hence the analysis of section~2 applies. In particular, the fractional part~$\kappa_{ff}\, \mod\,1$ is a good observable in any superconformal theory with a~$U(1)$ flavor symmetry.

\subsec{$R$-Symmetries}

Every three-dimensional~$\CN=2$ theory admits a supercurrent multiplet~$\CS_\mu$ that contains the supersymmetry current and the energy-momentum tensor, as well as other operators. A thorough discussion of supercurrents in three dimensions  can be found in~\DumitrescuIU. If the theory has a~$U(1)_R$ symmetry, the~$\CS$-multiplet can be improved to a multiplet~$\CR_\mu$, which satisfies
\eqn\rmult{\b D^\beta \CR_{\alpha\beta} = - 4 i \b D_\alpha \CJ^{(Z)}~, \qquad D^2 \CJ^{(Z)} = \b D^2 \CJ^{(Z)} = 0~.}
Here~$\CR_{\alpha\beta} = -2 \gamma^\mu_{\alpha\beta} \CR_\mu$ is the symmetric bi-spinor corresponding to~$\CR_\mu$. Note that~$\CJ^{(Z)}$ is a real linear multiplet, and hence~$\CR_\mu$ is also annihilated by~$D^2$ and~$\b D^2$. In components,
\eqn\rmultcomp{\eqalign{& \CR_\mu = j_\mu^{(R)} - i \theta S_\mu - i \b \theta \b S_\mu - (\theta\gamma^\nu\b \theta) \big(2 T_{\mu\nu} + i \ep_{\mu\nu\rho} \d^\rho J^{(Z)}\big) \cr
& \hskip25pt - i \theta \b \theta \big(2 j_\mu^{(Z)} + i \ep_{\mu\nu\rho} \d^\nu j^{(R)\rho} \big) + \cdots~, \cr
& \CJ^{(Z)} = J^{(Z)} - \half \theta \gamma^\mu S_\mu + \half \b \theta \gamma^\mu \b S_\mu + i \theta \b \theta T^\mu_\mu - (\theta \gamma^\mu \b \theta) j_\mu^{(Z)} + \cdots~,}}
where the ellipses denote terms that are determined by the lower components as in~\jcomp. Here~$j_\mu^{(R)}$ is the~$R$-current, $S_{\alpha\mu}$ is the supersymmetry current, $T_{\mu\nu}$ is the energy-momentum tensor, and~$j_\mu^{(Z)}$ is the current associated with the central charge in the supersymmetry algebra. The scalar~$J^{(Z)}$ gives rise to a string current~$i\ep_{\mu\nu\rho} \d^\rho J^{(Z)}$. All of these currents are conserved. Note that there are additional factors of~$i$ in~\rmultcomp\ compared to the formulas in~\DumitrescuIU, because we are working in Euclidean signature. (In Lorentzian signature the superfield~$\CR_\mu$ is real.)

The~$\CR$-multiplet is not unique. It can be changed by an improvement transformation,
\eqn\rimp{\eqalign{& \CR_{\alpha\beta}' = \CR_{\alpha\beta} - {t \over 2} \left([D_\alpha,\b D_\beta] + [D_\beta,\b D_\alpha]\right) \CJ~,\cr
& \CJ'^{(Z)} = \CJ^{(Z)} - {i t \over 2} \, \b D D \CJ~,}}
where~$\CJ$ is a flavor current and~$t$ is a real parameter. In components,
\eqn\impcop{\eqalign{&j_\mu'^{(R)} = j_\mu^{(R)} + t j_\mu~,\cr
& T_{\mu\nu}' =T_{\mu\nu} - {t \over 2} (\d_\mu \d_\nu - \delta_{\mu\nu}\d^2) J~,\cr
& J'^{(Z)} = J^{(Z)} + t K~,\cr
& j_\mu'^{(Z)} = j_\mu^{(Z)} - i t \ep_{\mu\nu\rho} \d^\nu j^\rho~.}}
Note that the~$R$-current~$j_\mu^{(R)}$ is shifted by the flavor current~$j_\mu$. If the theory is superconformal, it is possible to set~$\CJ^{(Z)}$ to zero by an improvement transformation, so that~$J^{(Z)}, T^\mu_\mu$, and~$j_\mu^{(Z)}$ are redundant operators.

We first consider the two-point functions of operators in the flavor current multiplet~$\CJ$ with operators in the~$\CR$-multiplet. They are parameterized by two dimensionless structure functions~$\hat \tau_{fr}$ and~$\kappa_{fr}$, where the subscript~$fr$ emphasizes the fact that we are considering mixed flavor-$R$ two-point functions:
\eqn\flavorRc{\eqalign{& \langle j_\mu(p) j_\nu^{(R)}(-p)\rangle = (p_\mu p_\nu - p^2 \delta_{\mu\nu}) {\hat \tau_{fr} \over 8 |p|} + \ep_{\mu\nu\rho} p^\rho {\kappa_{fr} \over 2 \pi}~,\cr
& \langle j_\mu(p) j_\nu^{(Z)}(-p)\rangle = (p_\mu p_\nu -p^2 \delta_{\mu\nu}) {\kappa_{fr} \over 2 \pi} - \ep_{\mu\nu\rho} p^\rho {|p| \hat \tau_{fr} \over 8}~,\cr
& \langle J(p) J^{(Z)}(-p)\rangle = {\kappa_{fr} \over 2\pi}~,\cr
& \langle K(p) J^{(Z)}(-p)\rangle = - {|p| \hat \tau_{fr} \over 8}~,\cr
& \langle J(p) T_{\mu\nu} (-p)\rangle = (p_\mu p_\nu - p^2 \delta_{\mu\nu} ) {\hat \tau_{fr} \over 16 |p|}~,\cr
& \langle K(p) T_{\mu\nu}(-p)\rangle = (p_\mu p_\nu - p^2 \delta_{\mu\nu}) {\kappa_{fr} \over 4 \pi}~.}}

Under an improvement transformation~\impcop, the structure functions shift as follows:
\eqn\impshift{\eqalign{
&\hat \tau'_{fr} = \hat \tau_{fr} + t \,\hat \tau_{ff}~,\cr
& \kappa'_{fr} =  \kappa_{fr} + t \, \kappa_{ff}~.}}
As explained above, in a superconformal theory there is a preferred~$\CR'_{\alpha\beta}$, whose corresponding~$\CJ'^{(Z)}$ is a redundant operator. Typically, it differs from a natural choice~$\CR_{\alpha\beta}$ in the UV by an improvement transformation~\rimp. In order to find the value of~$t$ that characterizes this improvement, we can use~\flavorRc\ and the fact that the operators in~$\CJ'^{(Z)}$ are redundant to conclude that~$\hat \tau'_{fr}$ must vanish~\BarnesBM. Alternatively, we can determine~$t$ by applying the~$F$-maximization principle, which was conjectured in~\refs{\JafferisUN,\JafferisZI} and proved in~\ClossetVG.

We will now discuss two-point functions of operators in the~$\CR$-multiplet. They are parameterized by four dimensionless structure functions~$\hat \tau_{rr}$, $\hat \tau_{zz}$, $\kappa_{rr}$, and~$\kappa_{zz}$,
\eqn\rtp{\eqalign{&\langle j_\mu^{(R)}(p) j_\nu^{(R)}(-p)\rangle = (p_\mu p_\nu - p^2 \delta_{\mu\nu}) {\hat \tau_{rr} \over 8 |p|} + \ep_{\mu\nu\rho} p^\rho {\kappa_{rr} \over 2 \pi}~,\cr
&\langle j_\mu^{(Z)}(p) j_\nu^{(Z)}(-p)\rangle = (p_\mu p_\nu - p^2 \delta_{\mu\nu}) {|p|\hat \tau_{zz} \over 8} + \ep_{\mu\nu\rho} p^\rho p^2 {\kappa_{zz} \over 2 \pi}~,\cr
& \langle j_\mu^{(Z)}(p) j_\nu^{(R)}(-p)\rangle = -(p_\mu p_\nu - p^2 \delta_{\mu\nu}) {\kappa_{zz} \over 2\pi} + \ep_{\mu\nu\rho} p^\rho {|p| \hat \tau_{zz} \over 8}~,\cr
& \langle J^{(Z)}(p) J^{(Z)}(-p) \rangle = {|p|\hat \tau_{zz} \over 8}~,\cr
& \langle J^{(Z)}(p) T_{\mu\nu} (-p) \rangle = - {\kappa_{zz} \over 4 \pi} (p_\mu p_\nu - p^2 \delta_{\mu\nu})~.}}
The two-point function~$\langle T_{\mu\nu}(p)T_{\rho\lambda}(-p)\rangle$ is given by~\emtensor\ with
\eqn\susytt{\tau_g = {\hat \tau_{rr} + 2 \hat \tau_{zz} \over 32}~, \qquad \tau_g' = -{\hat \tau_{rr} + \hat \tau_{zz} \over32}~, \qquad \kappa_g = 12 \left(\kappa_{rr} + \kappa_{zz}\right)~.}
The subscripts~$rr$ and~$zz$ are associated with two-point functions of the currents~$j_\mu^{(R)}$ and~$j_\mu^{(Z)}$. Note that~$\tau_g +\tau_g' =  {{\hat \tau}_{zz} \over 32}$, which is non-negative and vanishes in a superconformal theory. As before, an improvement transformation~\rimp\ shifts the structure functions,
\eqn\rrimp{\eqalign{& \hat \tau_{rr}' = \hat \tau_{rr} + 2t \,  \hat \tau_{fr} + t^2 \, \hat \tau_{ff}~,\cr
& \hat \tau_{zz}' = \hat \tau_{zz} - 2t \, \hat \tau_{fr} - t^2 \, \hat \tau_{ff}~,\cr
& \kappa_{rr}' = \kappa_{rr} + 2 t \, \kappa_{fr} + t^2\, \kappa_{ff}~,\cr
& \kappa_{zz}' = \kappa_{zz} - 2 t \, \kappa_{fr} - t^2 \, \kappa_{ff}~.}}
Note that~$\tau_g'$ and~$\kappa_g$ in~\susytt\ are invariant under these shifts.

In a superconformal theory, the operators~$J^{(Z)}$, $T^\mu_\mu$, and~$j_\mu^{(Z)}$ are redundant. However, we see from~\flavorRc\ and~\rtp\ that they give rise to contact terms, which are parameterized by~$\kappa_{fr}$ and~$\kappa_{zz}$. These contact terms violate conformal invariance. Unless~$\kappa_{fr}$ and~$\kappa_{zz}$ are properly quantized, they cannot be set to zero by a local counterterm without violating the quantization conditions for Chern-Simons counterterms explained in the introduction. This leads to a new anomaly, which will be discussed in section~5.

\subsec{Background Supergravity Fields}

In order to get a better understanding of the contact terms discussed in the previous subsection, we couple the~$\CR$-multiplet to background supergravity fields. (See appendix~B for relevant aspects of~$\CN=2$ supergravity.) To linear order, the~$\CR$-multiplet couples to the linearized metric superfield~$\CH_\mu$. In Wess-Zumino gauge,
\eqn\hcomp{\CH_\mu = \half \left(\theta \gamma^\nu \b \theta\right) \left(h_{\mu\nu} - i B_{\mu\nu}\right) - {1 \over 2} \theta \b \theta C_\mu- {i \over 2} \thetasq \b \theta \b \psi_\mu +{i \over 2} \b \thetasq \theta \psi_\mu + \half \thetasq \b \thetasq \left(A_\mu-V_\mu\right)~.}
Here~$h_{\mu\nu}$ is the linearized metric, so that~$g_{\mu\nu} = \delta_{\mu\nu} + 2 h_{\mu\nu}$. The vectors~$C_\mu$ and~$A_\mu$ are Abelian gauge fields, and~$B_{\mu\nu}$ is a two-form gauge field. It will be convenient to define the following field strengths,
\eqn\fsdef{\eqalign{& V_\mu = - \ep_{\mu\nu\rho} \d^\nu C^\rho~, \qquad \d^\mu V_\mu = 0~,\cr
& H = {1 \over 2} \ep_{\mu\nu\rho} \d^\mu B^{\nu\rho}~.}}
Despite several unfamiliar factors of~$i$ in~\hcomp\ that arise in Euclidean signature, the fields~$V_\mu$ and~$H$ are naturally real. Below, we will encounter situations with imaginary~$H$, see also~\refs{\FestucciaWS,\ClossetVG}.

If the theory is superconformal, we can reduce the~$\CR$-multiplet to a smaller supercurrent. Consequently, the linearized metric superfield~$\CH_\mu$ enjoys more gauge freedom, which allows us to set~$B_{\mu\nu}$ and~$A_\mu - \half V_\mu$ to zero. The combination~$A_\mu-{3\over 2}V_\mu$ remains and transforms like an Abelian gauge field.

Using~$\CH_\mu$, we can construct three Chern-Simons terms (see appendix~B), which capture the contact terms described in the previous subsection. As we saw there, not all of them are conformally invariant.

\medskip

\item{$\bullet$} {\it Gravitational Chern-Simons Term:}
\eqn\lcs{\eqalign{{\scr L}_g = & {\kappa_g \over 192 \pi} \Big( i\ep^{\mu\nu\rho} \Tr \big(\omega_\mu \d_\nu \omega_\rho + {2 \over 3} \omega_\mu \omega_\nu \omega_\rho\big)\cr
& + 4 i \ep^{\mu\nu\rho} \big(A_\mu - {3 \over 2} V_\mu\big) \d_\nu \big(A_\rho - {3 \over 2} V_\rho\big) + \left({\rm fermions}\right)\Big)~.}}
We see that the~$\CN=2$ completion of the gravitational Chern-Simons term~\lcsdef\ also involves a Chern-Simons term for~$A_\mu -{3 \over 2} V_\mu$. Like the flavor-flavor term~\ffcs, the gravitational Chern-Simons term~\lcs\ is conformally invariant. It was previously studied in the context of conformal~$\CN=2$ supergravity~\RocekBK, see also~\refs{\AchucarroVZ,\AchucarroGM}.

\medskip

\item{$\bullet$} {\it $Z$-$Z$ Chern-Simons Term:}
\eqn\ggcs{{\scr L}_{zz} = -{ \kappa_{zz} \over 4 \pi}\left(i \ep^{\mu\nu\rho} \big(A_\mu - \half V_\mu \big) \d_\nu \big( A_\rho - \half V_\rho\big) + \half H R + \cdots + \left({\rm fermions}\right)\right)~.}
Here the ellipsis denotes higher-order terms in the bosonic fields, which go beyond linearized supergravity. The presence of the Ricci scalar~$R$ and the fields~$H$, $A_\mu - \half V_\mu$ implies that~\ggcs\ is not conformally invariant.

\medskip

\item{$\bullet$} {\it Flavor-R Chern-Simons Term:}
\eqn\frcs{{\scr L}_{fr} = -{\kappa_{fr} \over 2\pi} \left(i \ep^{\mu\nu\rho} a_\mu\d_\nu \big(A_\rho - \half V_\rho\big) + {1 \over 4} \sigma R - D H + \cdots + \left({\rm fermions}\right)\right)~.}
The meaning of the ellipsis is as in~\ggcs.  Again, the presence of~$R$,~$H$, and~$A_\mu - {1\over 2} V_\mu$ shows that this term is not conformally invariant. The relative sign between the Chern-Simons terms~\ffcs\ and~\frcs\ is due to the different couplings of flavor and~$R$-currents to their respective background gauge fields.
\medskip

\noindent Note that both~\lcs\ and~\ggcs\ give rise to a Chern-Simons term for~$A_\mu$. Its overall coefficient is~$\kappa_{rr} = {\kappa_g \over 12} - \kappa_{zz}$, in accord with~\susytt.

It is straightforward to adapt the discussion of section~2 to these Chern-Simons terms. Their coefficients can be modified by shifting the Lagrangian by appropriate counterterms, whose coefficients are quantized according to the periodicity of the global symmetries. Instead of stating the precise quantization conditions, we will abuse the language and say that the fractional parts of these coefficients are physical, while their integer parts are scheme dependent.

\newsec{A New Anomaly}

In the previous section, we have discussed four Chern-Simons terms in the background fields: the flavor-flavor term~\ffcs, the gravitational term~\lcs, the~$Z$-$Z$ term~\ggcs, and the flavor-$R$ term~\frcs. They correspond to certain contact terms in two-point functions of operators in the flavor current~$\CJ$ and the~$\CR$-multiplet. As we saw above, the flavor-flavor and the gravitational Chern-Simons terms are superconformal, while the~$Z$-$Z$ term and the flavor-$R$ term are not. The latter give rise to non-conformal contact terms proportional to~$\kappa_{zz}$ and~$\kappa_{fr}$.

The integer parts of~$\kappa_{zz}$ and~$\kappa_{fr}$ can be changed by adding appropriate Chern-Simons counterterms, but the fractional parts are physical and cannot be removed. This leads to an interesting puzzle: if~$\kappa_{zz}$ or~$\kappa_{fr}$ have non-vanishing fractional parts in a superconformal theory, they give rise to non-conformal contact terms. This is similar to the conformal anomaly in two dimensions, where the redundant operator~$T_\mu^\mu$ has nonzero contact terms. However, in two-dimensions the non-conformal contact terms arise from correlation functions of the conserved energy-momentum tensor at separated points, and hence they cannot be removed by a local counterterm. In our case, the anomaly is a bit more subtle.

An anomaly arises whenever we are unable to impose several physical requirements at the same time. Although the anomaly implies that we must sacrifice one of these requirements, we can often choose which one to give up. In our situation we would like to impose supersymmetry, conformal invariance, and compactness of the global symmetries, including the~$R$-symmetry. Moreover, we would like to couple the global symmetries to arbitrary background gauge fields in a fully gauge-invariant way. As we saw above, this implies that the corresponding Chern-Simons counterterms must have integer coefficients.\foot{Here we will abuse the language and attribute the quantization of these coefficients to invariance under large gauge transformations. As we reviewed in the introduction, a more careful construction requires a choice of auxiliary four-manifold. The quantization follows by demanding that our answers do not depend on that choice.} If the fractional part of~$\kappa_{zz}$ or $\kappa_{fr}$ is nonzero, we cannot satisfy all of these requirements, and hence there is an anomaly. In this case we have the following options:
\item{1.)} {\it We can sacrifice supersymmetry.}  Then we can shift the Lagrangian by non-supersymmetric counterterms that remove the non-conformal terms in~\ggcs\ and~\frcs\ and restore conformal invariance. Note that these counterterms are gauge invariant.
\item{2.)} {\it We can sacrifice conformal invariance.}  Then there is no need to add any counterterm. The correlation functions at separated points are superconformal, while the contact terms are supersymmetric but not conformal.
\item{3.)} {\it We can sacrifice invariance under large gauge transformations.} Now we can shift the Lagrangian by supersymmetric Chern-Simons counterterms with fractional coefficients to restore conformal invariance. These counterterms are not invariant under large gauge transformations, if the background gauge fields are topologically non-trivial.

\medskip

The third option is the most conservative, since we retain both supersymmetry and conformal invariance. If the background gauge fields are topologically non-trivial, the partition function is multiplied by a phase under large background gauge transformations. In order to obtain a well-defined answer, we need to specify additional geometric data.\foot{More precisely, the phase of the partition function depends on the choice of auxiliary four-manifold, which is the additional data needed to obtain a well-defined answer.} By measuring the change in the phase of the partition function as we vary this data, we can extract the fractional parts of~$\kappa_{zz}$ and~$\kappa_{fr}$. Therefore, these observables are not lost, even if we set the corresponding contact terms to zero by a counterterm.

This discussion is similar to the framing anomaly of~\WittenHF.  There, a Lorentz Chern-Simons term for the frame bundle is added with fractional coefficient, in order to make the theory topologically invariant.  This introduces a dependence on the trivialization of the frame bundle.  In our case the requirement of topological invariance is replaced with superconformal invariance and we sacrifice invariance under large gauge transformations rather than invariance under a change of framing.

Finally, we would like to point out that the anomaly described above has important consequences if the theory is placed on a curved manifold~\ClossetVG. For some configurations of the background fields, the partition function is not consistent with conformal invariance and even unitarity.

\newsec{A Perturbative Example: SQED with a Chern-Simons Term}

Consider~$\CN=2$ SQED with a level~$k$ Chern-Simons term for the dynamical~$U(1)_v$ gauge field and~$N_f$ flavor pairs~$Q_i, \t Q_{\t i}$ that carry charge~$\pm 1$ under~$U(1)_v$. The theory also has a global~$U(1)_a$ flavor symmetry under which~$Q_i, \t Q_{\t i}$ all carry charge~$+1$. Here~$v$ and~$a$ stand for `vector' and `axial' respectively. The Euclidean flat-space Lagrangian takes the form
\eqn\exflat{{\scr L} = - \int d^4 \theta \, \left(\b Q_i e^{2 \hat \CV_v} Q_i + {\b {\t Q}}_{\t i} e^{- 2 \hat \CV_v} \t Q_{\t i} -{1 \over e^2} \hat \Sigma_v^2 + {k \over 2 \pi} \hat \CV_v \hat \Sigma_v \right)~,}
where~$e$ is the gauge coupling and~$\hat \CV_v$ denotes the dynamical~$U(1)_v$ gauge field. (The hat emphasizes the fact that it is dynamical.) Note that the theory is invariant under charge conjugation, which maps~$\hat \CV_v \rightarrow - \hat \CV_v$ and~$Q_i \leftrightarrow \t Q_{\t i}$.  This symmetry prevents mixing of the axial current with the topological current, so that some of the subtleties discussed in section~3 are absent in this theory.

The Chern-Simons term leads to a mass for the dynamical gauge multiplet,
\eqn\photm{M = {k e^2 \over 2 \pi}~.}
This mass is the crossover scale from the free UV theory to a non-trivial CFT labeled by~$k$ and~$N_f$ in the IR. We will analyze this theory in perturbation theory for~$k \gg 1$. In particular, we will study the contact terms of the axial current,
\eqn\axcurr{\CJ = |Q_i|^2 + |\t Q_{\t i}|^2~,}
and the~$\CR$-multiplet,
\eqn\rmultex{\eqalign{& \CR_{\alpha\beta} = {2 \over e^2} \big(D_\alpha \hat \Sigma_v \b D_\beta \hat \Sigma_v + D_\beta \hat \Sigma_v \b D_\alpha \hat \Sigma_v\big) + \CR^m_{\alpha\beta}~,\cr
& \CJ^{(Z)} = {i \over 4 e^2} \b D D \big(\hat \Sigma_v^2\big)~.}}
Here~$\CR^m_{\alpha\beta}$ is associated with the matter fields and assigns canonical dimensions to~$Q_i, \t Q_{\t i}\,$. In the IR, the~$\CR$-multiplet flows to a superconformal multiplet, up to an improvement by the axial current~$\CJ$. Therefore, at long distances~$\CJ^{(Z)}$ is proportional to~$i \b D D \CJ$.

\medskip

\ifig\fvone{Feynman diagrams for flavor-flavor.  The solid dots denote the appropriate operator insertions.  The dashed and solid lines represent scalar and fermion matter.  The double line denotes the scalar and the auxiliary field in the vector multiplet, while the zigzag line represents the gaugino.}
{\epsfxsize0.95in\epsfbox{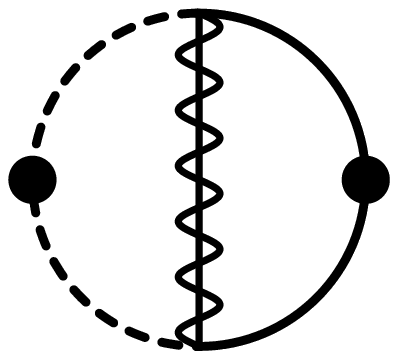} \hskip25pt
\hskip25pt \epsfxsize2.2in \epsfbox{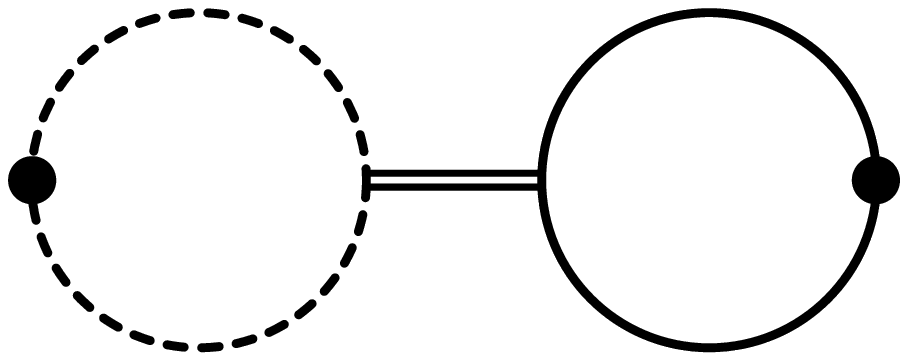}}

\medskip

We begin by computing the flavor-flavor contact term~$\kappa_{ff, {\rm CFT}}$ in the two-point function of the axial current~\axcurr, by flowing from the free UV theory to the CFT in the IR. Using~\threefun, we see that it suffices to compute the correlation function~$\langle  J(p) K(-p)\rangle$ at small momentum~$p^2 \rightarrow 0$. In a conformal field theory, the correlator~$\langle J(x) K(0)\rangle$ vanishes at separated points, and hence we must obtain a pure contact term. More explicitly, we have
\eqn\jkcomp{J = |q_i|^2 + |\t q_{\t i}|^2~, \qquad K = -i \b \psi_i \psi_i - i \b {\t \psi}_{\t i} \t \psi_{\t i}~.}
There are two diagrams at leading order in~$1\over k$, displayed in~\fvone. The first diagram, with the intermediate gaugino, is paired with a seagull diagram, which ensures that we obtain a pure contact term. The second diagram vanishes by charge conjugation. Evaluating these diagrams, we find
\eqn\kjct{\lim_{p^2 \rightarrow 0}\,  \langle J(p) K(-p)\rangle = {\pi N_f \over 8 k} +\CO\left({1\over k^3}\right)~,}
and hence
\eqn\exffsph{\kappa_{ff, {\rm CFT}} = {\pi^2 N_f \over 4 k} + \CO\left({1\over k^3}\right)~.}

\medskip

\ifig\ftwo{Feynman diagrams for flavor-gravity. See~\fvone\ for an explanation of the diagrammatic rules.}
{\epsfxsize=1.7in \epsfbox{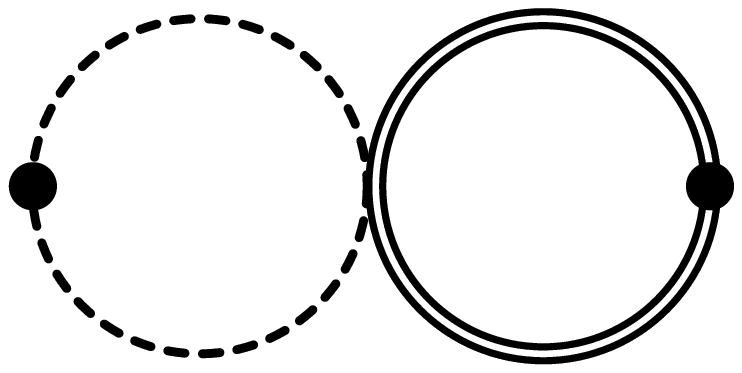} \hskip10pt \epsfxsize=1.5in \epsfbox{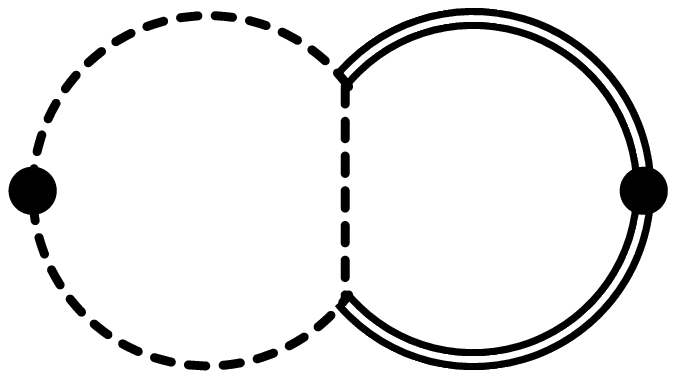} \hskip10pt  \hskip10pt
\epsfxsize=0.95in \epsfbox{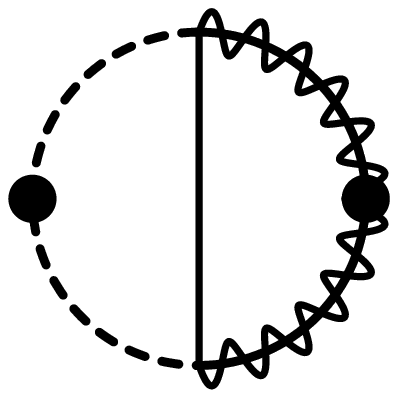} }

\medskip

We similarly compute the flavor-$R$ contact term~$\kappa_{fr, {\rm CFT}}$ by flowing into the CFT from the free UV theory. It follows from~\flavorRc\ that it can be determined by computing the two-point function~$\langle J(p) J^{(Z)}(-p)\rangle$ at small momentum~$p^2 \rightarrow 0$. Using~\rmultex, we find
\eqn\jzop{J^{(Z)} = -{1 \over e^2} \left(\hat \sigma_v \hat D_v - {i \over 2} {\b {\hat  \lambda}}_v \hat \lambda_v\right)~.}
Since~$\CJ^{(Z)}$ is proportional to~$i \b D D \CJ$ at low energies, the operator~$J^{(Z)}$ flows to an operator proportional to~$K$. The coefficient is determined by the mixing of the~$R$-symmetry with the axial current~$\CJ$, which occurs at order~${1 \over k^2}$. Since~$\langle J(x) K(0)\rangle$ vanishes at separated points, the two-point function of~$J$ and~$J^{(Z)}$ must be a pure contact term. Unlike the flavor-flavor case, several diagrams contribute to this correlator at order $1\over k$ (\ftwo). Each diagram gives rise to a term proportional to~$1 \over |p|$. However, these contributions cancel, and we find a pure contact term,
\eqn\jzjct{\lim_{p^2 \rightarrow 0} \langle J(p) J^{(Z)} (-p)\rangle = -{N_f \over 4 \pi k} +\CO\left({1\over k^3}\right)~,}
so that
\eqn\exfrsph{\kappa_{fr, {\rm CFT}} =- {N_f \over 2k} + \CO\left({1\over k^3}\right) ~.}
Since this value is fractional, it implies the presence of the anomaly discussed in the previous section.

We have computed~$\kappa_{ff, {\rm CFT}}$ and~$\kappa_{fr, {\rm CFT}}$ by flowing into the CFT from the free UV theory. It is instructive to follow the discussion in subsection~3.4 and further deform the theory by a real mass~$m \ll M$. In order to preserve charge conjugation, we assign the same real mass~$m$ to all flavors~$Q_i, \t Q_{\t i}$. This deformation leads to a gap in the IR. Even though a topological theory with Lagrangian proportional to~$i \ep^{\mu\nu\rho}\hat v_\mu\partial_\nu \hat v_\rho$ can remain, it does not mix with~$\CJ$ or~$\CR_{\alpha\beta}$ because of charge conjugation. Therefore, the contact terms~$\kappa_{ff}$ and~$\kappa_{fr}$ must be properly quantized in the IR. (Since the matter fields in this example have half-integer~$R$-charges, this means that~$\kappa_{fr}$ should be a half-integer.)

For the axial current, we have
\eqn\taup{\hat \tau_{ff} \approx \cases{
2N_f & $ \quad p^2\gg M^2$ \cr
\hat \tau_{ff, {\rm CFT}}= 2 N_f-\CO\left({1\over k^2}\right) & \quad $m^2\ll p^2 \ll M^2$\cr
0 & $ \quad p^2\ll m^2$}}
The fact that~$\hat \tau_{ff} = 0$ in the IR follows from the fact that the theory is gapped.  Similarly,
\eqn\kappap{\kappa_{ff} \approx \cases{
0 & $\quad p^2\gg M^2$ \cr
\kappa_{ff, {\rm CFT}} ={\pi^2 N_f\over 4k} + \CO\left({1\over k^3}\right) &\quad  $m^2\ll p^2 \ll M^2$\cr
-N_f \sign (m) & $ \quad p^2\ll m^2$}}
Note that parity, which acts as~$k \to -k$,~$m \to -m$,~$\kappa_{ff} \to -\kappa_{ff}$, with~$\hat \tau_{ff}$ invariant, is a symmetry of~\taup\ and~\kappap.

For the two-point function of the axial current and the~$\CR$-multiplet, we find
\eqn\taufr{\hat \tau_{fr}\approx \cases{
0 & $\quad p^2\gg M^2$ \cr
\hat \tau_{fr, {\rm CFT}}= \CO\left({1\over k^2}\right) & \quad $m^2\ll p^2 \ll M^2$\cr
0 & $ \quad p^2\ll m^2$}}
Here~$\hat \tau_{fr, {\rm CFT}}$ measures the mixing of the axial current with the UV~$\CR$-multiplet~\rmultex. For the superconformal~$\CR$-multiplet of the CFT, we would have obtained~$\hat \tau_{fr, {\rm CFT}} = 0$, as explained after~\impshift. Similarly,
\eqn\kappapfr{\kappa_{fr} \approx \cases{
0 & $ \quad p^2\gg M^2$ \cr
\kappa_{fr, {\rm CFT}}=-{N_f \over 2 k}  + \CO\left({1\over k^3}\right) & \quad $m^2\ll p^2 \ll M^2$\cr
{N_f \over 2} \sign (m) & \quad  $p^2\ll m^2$}}
As before,~\taufr\ and~\kappapfr\ transform appropriately under parity.

Let us examine the flow from the CFT to the IR in more detail, taking the UV crossover scale~$M\to \infty$. In the CFT, the operator~$J^{(Z)}$ is redundant, up to~$\CO\left(1\over k^2\right)$ corrections due to the mixing with the axial current. Once the CFT is deformed by the real mass~$m$, we find that
\eqn\JZJm{J^{(Z)} = m J +\CO\left({1\over e^2}, {1\over k^2}\right)~,}
where~$J$ is the bottom component of the axial current~\axcurr, which is given by~\jkcomp.  (As always, the operator equation~\JZJm\ holds at separated points.)  Substituting into~\flavorRc, we find that
\eqn\kappalow{{\kappa_{fr}\over 2 \pi} = {\kappa_{fr, {\rm CFT}} \over 2 \pi}+m \langle J(p) J(-p) \rangle  + \CO\left({1\over e^2}, {1\over k^2}\right) = {\kappa_{fr, {\rm CFT}} \over 2 \pi} + {m \over 8 |p| } \hat \tau_{ff} + \CO\left({1\over e^2}, {1\over k^2}\right)~.}
Here it is important that the two-point function of~$J$ does not have a contact term in the CFT. Explicitly computing~$\hat \tau_{ff}$, we find that
\eqn\taulow{\hat \tau_{ff}=\cases{2 N_f - \CO\left({1 \over k^2}\right) &$\quad p^2 \gg m^2 $ \cr
{|p|\over |m|}{2 N_f \over \pi} \left(1 + {1 \over k} \sign (m) \right)+\CO\left({1 \over k^2}\right) &\quad $p^2\ll m^2$}}
This is consistent with~\kappapfr\ and~\kappalow.

\newsec{Checks of Dualities}

In this section we examine dual pairs of three-dimensional~$\CN=2$ theories, which are conjectured to flow to the same IR fixed point. In this case, the various contact terms discussed above, computed on either side of the duality, should match.

First, as in~\refs{\KapustinHA\KapustinXQ\KapustinMH\WillettGP-\BeniniMF}, the three-sphere partition functions of the two theories should match, up to the contribution of Chern-Simons counterterms in the background fields.  Denote their coefficients by~$\delta \kappa$.

Second, as in the parity anomaly matching condition discussed in~\AharonyBX, the fractional parts of these contact terms are intrinsic to the theories.  Therefore, the Chern-Simons counterterms that are needed for the duality must be properly quantized.  This provides a new non-trivial test of the duality.

Finally, these counterterms can often be determined independently.  Whenever different pairs of dual theories are related by renormalization group flows, the counterterms for these pairs are similarly related.  In particular, given the properly quantized Chern-Simons counterterms that are needed for one dual pair, we can determine them for other related pairs by a one-loop computation in flat space.  This constitutes an additional check of the duality.

In this section we demonstrate this matching for~$\CN=2$ supersymmetric level-rank duality and Giveon-Kutasov duality~\GiveonZN. We compute some of the relative Chern-Simons counterterms, both in flat space and using the three-sphere partition function, and verify that they are properly quantized.

\subsec{Level-Rank Duality}

Consider an~$\CN=2$ supersymmetric~$U(n)$ gauge theory with a level $k$ Chern-Simons term. We will call this the `electric' theory and denote it by~$U(n)_k$. In terms of the $SU(n)$ and $U(1)$ subgroups, this theory is equivalent to $\left(SU(n)_k \times U(1)_{nk}\right)/ \Z_n$, where we have used the conventional normalization for Abelian gauge fields.  This theory flows to a purely topological~$U(n)$ Chern-Simons theory with shifted levels, denoted by~$U(n)^{top}_{\sign (k) \left(|k| - n\right), \, k n}$. The first subscript specifies the level of the~$SU(n)$ subgroup, which is shifted by integrating out the charged, massive gauginos (recall that their mass has the same sign as the level~$k$), and the second subscript denotes the level of the~$U(1)$ subgroup, which is not shifted.

The dual `magnetic' theory is a supersymmetric~$U(|k| - n)_{-k}$ Yang-Mills Chern-Simons theory. It flows to the purely topological theory~$U(|k| - n)^{top}_{-\sign (k) n, - k (|k| - n)}$~. This theory is related to the other topological theory described above by conventional level-rank duality for unitary gauge groups~\NaculichNC.\foot{The authors of~\NaculichNC\ restricted $n$ to be odd and~$k$ to be even.  This restriction is unnecessary on spin manifolds.  Furthermore, we reversed the orientation on the magnetic side.}

These theories have two Abelian symmetries: a~$U(1)_R$ symmetry under which all gauginos have charge~$+1$, and a topological symmetry~$U(1)_J$. The topological symmetry corresponds to the current~$j_\mu = {i \over 2 \pi} \ep_{\mu\nu\rho}\Tr F^{\nu\rho}$ on the electric side, and to~$j_\mu = - {i \over 2 \pi} \ep_{\mu\nu\rho} \Tr F^{\nu\rho}$ on the magnetic side.

We can integrate out the gauginos to obtain the contact term~$\kappa_{rr}$ in the two-point function~\rtp\ of the~$R$-current. On the electric side, we find~$\kappa_{rr, e} = - \half \sign (k) n^2$, and on the magnetic side we have~$\kappa_{rr, m} = \half \sign (k) (|k| - n)^2$. We must therefore add a counterterm
\eqn\deltakappaR{\delta\kappa_{rr} = - \half \sign (k) \left( (|k|-n)^2 + n^2\right)~,}
to the magnetic theory. Taking into account possible half-integer counterterms that must be added on either side of the duality because of the parity anomaly, what remains of the relative counterterm~\deltakappaR\ is always an integer.

In order to compute the contact term associated with~$U(1)_J$, we follow the discussion in subsection~3.2 and integrate out the dynamical gauge fields to find the effective theory for the corresponding background gauge field. In the electric theory, this leads to~$\kappa_{JJ, e} = - {n \over k}$, and in the magnetic theory we find~$\kappa_{JJ, m} = {|k| - n \over k}$. Hence we need to add an integer Chern-Simons counterterm to the magnetic theory,
\eqn\deltakappaJ{\delta\kappa_{JJ}= -\sign (k)~.}

\subsec{Giveon-Kutasov Duality}

Consider the duality of Giveon and Kutasov~\GiveonZN. The electric theory consists of a~$U(n)_k$ Chern-Simons theory with~$N_f$ pairs~$Q_i, \t Q_{\t i}$ of quarks in the fundamental and the anti-fundamental representation of~$U(n)$. The global symmetry group is $SU(N_f)\times SU(N_f)\times U(1)_A \times U(1)_J\times U(1)_R$. The quantum numbers of the fundamental fields are given by
\eqn\GKe{
\vbox{\offinterlineskip\tabskip=0pt
\halign{\strut\vrule#
&~$#$~\hfil\vrule
&~$#$~\hfil\vrule
&~$#$~\hfil
&~$#$~\hfil
&~$#$\hfil
&~$#$\hfil
&~$#$\hfil
&\vrule#
\cr
\noalign{\hrule}
&{\rm Fields} &  U(n)_k & SU(N_f)&   SU(N_f)& U(1)_A & U(1)_J & U(1)_R &\cr
\noalign{\hrule}
&  {Q}         & \; {\bf  \square}     & \; {\bf \square}    & \; {\bf 1} & \quad 1  &    \quad  0 &  \quad  \half   &\cr
&  \tilde Q                & \; {\bf \b {\square}}     & \; {\bf 1}   & \; {\bf\b {\square}}   & \quad 1  &    \quad  0 &   \quad \half   &\cr
\noalign{\hrule}
}}}

The magnetic dual is given by a~$U(\tilde n= N_f + |k|-n)_{-k}$ Chern-Simons theory. It contains~$N_f$ pairs~$q^i, \t q_{\t i}$ of dual quarks and~$N_f^2$ singlets~${M_i}^{\t i}$, which interact through a superpotential~$W= q^i {M_i}^{\t i} \t q_{\t i}$. The quantum numbers in the magnetic theory are given by
\eqn\GKm{
\vbox{\offinterlineskip\tabskip=0pt
\halign{\strut\vrule#
&~$#$~\hfil\vrule
&~$#$~\hfil\vrule
&~$#$~\hfil
&~$#$~\hfil
&~$#$\hfil
&~$#$\hfil
&~$#$\hfil
&\vrule#
\cr
\noalign{\hrule}
&{\rm Fields} &  U(\t n)_{-k} & SU(N_f)&   SU(N_f)& U(1)_A & U(1)_J & U(1)_R &\cr
\noalign{\hrule}
&  {q}         & \; {\bf {\square}}     & \; {\bf \b{ \square}}    &\; {\bf  1} & \quad -1  &    \quad  0 &  \quad  \half   &\cr
& \tilde q                & \; {\bf \b { \square} }    & \; {\bf 1}   & \; {\bf\square}   & \quad -1  &    \quad  0 &   \quad \half   &\cr
& M               & \; {\bf  1}     & \; {\bf \square}   & \; {\bf \b {\square}}   & \quad 2  &    \quad  0 &  \quad  1   &\cr}
\hrule}}

As before, the topological symmetry~$U(1)_J$ corresponds to~$j_\mu = {i \over 2\pi} \ep_{\mu\nu\rho} \Tr F^{\nu\rho}$ on the electric side, and to~$j_\mu = -{i  \over 2 \pi} \ep_{\mu\nu\rho}\Tr F^{\nu\rho}$ on the magnetic side. Note that none of the fundamental fields are charged under~$U(1)_J$.

This duality requires the following Chern-Simons counterterms for the Abelian symmetries, which must be added to the magnetic theory:\foot{Similar counterterms are required for the~$SU(N_f)\times SU(N_f)$ flavor symmetry~\refs{\KapustinXQ\KapustinMH\WillettGP-\BeniniMF}.}
\eqn\deltakappGK{\eqalign{
& \delta\kappa_{AA} = -\sign (k)  N_f(N_f-|k|)~, \cr
& \delta\kappa_{JJ} = -\sign (k)~,\cr
& \delta \kappa_{A r}=  \half \sign (k) N_f(N_f + |k| - 2 n)~, \cr
&\delta\kappa_{rr} = -{1 \over 4} \sign (k) \left(2 k^2 - 4 |k| n + 3 |k| N_f + 4 n^2 - 4 n N_f+ N_f^2\right)~.}}
This was derived in~\BeniniMF\ by flowing into Giveon-Kutasov duality from Aharony duality~\AharonyGP\ via a real mass deformation.\foot{The~$R$-symmetry used in~\BeniniMF\ assigns~$R$-charge~$0$ to the electric quarks~$Q_i, \t Q_{\t i}\,$. Therefore, our results for~$\delta\kappa_{Ar}$ and~$\delta\kappa_{rr}$ differ from those of~\BeniniMF\ by improvements~\impshift\ and~\rrimp.} Note that these Chern-Simons counterterms are properly quantized:~$\delta\kappa_{AA}$ and~$\delta\kappa_{JJ}$ are integers, while $\delta\kappa_{Ar}$ is half-integer and~$\delta\kappa_{rr}$ is quantized in units of~$1 \over 4$. This is due to the presence of fields with~$R$-charge~$\half$.

We can also understand~\deltakappGK\ by flowing out of Giveon-Kutasov duality to a pair of purely topological theories. If we give a real mass to all electric quarks, with its sign opposite to that of the Chern-Simons level~$k$, we flow to a~$U(n)_{k + \sign (k) N_f}$ theory without matter. The corresponding deformation of the magnetic theory flows to~$U(|k| - n)_{-(k + \sign (k) N_f)}$. Level-rank duality between these two theories without matter was discussed in the previous subsection. Given the counterterms~\deltakappaR\ and~\deltakappaJ\ that are needed for this duality and accounting for the
Chern-Simons terms generated by the mass deformation, we reproduce~\deltakappGK.

\subsec{Matching the Three-Sphere Partition Function}

As explained in~\ClossetVG, we can read off the contact terms~$\kappa_{ff}$ and~$\kappa_{fr}$ from the dependence of the free energy~$F_{S^3}$ on a unit three-sphere on the real mass parameter~$m$ associated with the flavor symmetry:
\eqn\review{\kappa_{ff} = -{1 \over 2 \pi} {\d^2 \over \d m^2} \Im F_{S^3}   \bigg|_{m = 0}~, \qquad \kappa_{fr} = {1 \over 2 \pi} {\d \over \d m} \Re F_{S^3}  \bigg|_{m= 0}~.}

We can use this to rederive some of the relative Chern-Simons counterterms in~\deltakappGK. Let us denote by~$m$ and~$\xi$ the real mass parameters corresponding to~$U(1)_A$ and~$U(1)_J$. (Equivalently, $\xi$ is a Fayet-Iliopoulos term for the dynamical gauge fields.) Using the results of~\vdB, it was shown in~\WillettGP\ that the difference between the three-sphere partition functions of the electric and the magnetic theories requires a counterterm
\eqn\wyct{\delta F_{S^3} = \sign (k) \left(\pi i N_f (N_f - |k|) m^2 + \pi i \xi^2 + \pi N_f (N_f + |k| - 2 n) m\right)+ \cdots~.}
where the ellipsis represents terms that are independent of $m$ and $\xi$. (Our conventions for the Chern-Simons level~$k$ differ from those of~\WillettGP\ by a sign.) An analogous result was obtained in~\BeniniMF\ for a different choice of~$R$-symmetry. Using~\review, we find the same values for~$\delta\kappa_{AA}$, $\delta\kappa_{JJ}$, and~$\delta\kappa_{Ar}$ as in~\deltakappGK. Note that the counterterm~\wyct\ does not just affect the phase of the partition function, because the term linear in~$m$ is real.

Many other dualities have been shown to require relative Chern-Simons counterterms~\refs{\KapustinHA\KapustinXQ\KapustinMH\JafferisNS\WillettGP\BeniniMF-\DimofteJU}. It would be interesting to repeat the preceding analysis in these examples.

\vskip 1cm

\noindent {\bf Acknowledgments:}
We would like to thank O.~Aharony, S.~Cremonesi, D.~Freed, D.~Gaiotto, D.~Jafferis, A.~Kapustin, I.~Klebanov, J.~Maldacena, A.~Schwimmer, B.~Willett, E.~Witten, and~I.~Yaakov for many useful discussions. CC is a Feinberg postdoctoral fellow at the Weizmann Institute of Science. The work of TD was supported in part by a DOE Fellowship in High Energy Theory and a Centennial Fellowship from Princeton University. The work of GF was supported in part by NSF grant PHY-0969448. TD and GF would like to thank the Weizmann Institute of Science for its kind hospitality during the completion of this project.  ZK was supported by NSF grant PHY-0969448 and a research grant from Peter and Patricia Gruber Awards, as well as by the Israel Science Foundation under grant number~884/11.  The work of NS was supported in part by DOE grant DE-FG02-90ER40542. ZK and NS would like to thank the United States-Israel Binational Science Foundation (BSF) for support under grant number~2010/629.
Any opinions, findings, and conclusions or recommendations expressed in this
material are those of the authors and do not necessarily reflect the views of the funding agencies.

\appendix{A}{Free Massive Theories}

Consider a complex scalar field~$\phi$ of mass~$m$,
\eqn\philag{{\scr L} = |\d_\mu \phi|^2 + m^2 |\phi|^2~.}
This theory is invariant under parity and has a~$U(1)$ flavor symmetry under which~$\phi$ has charge~$+1$. The corresponding current is given by
\eqn\phicur{j_\mu = i\left(\b \phi \d_\mu \phi - \phi \d_\mu \b \phi\right)~.}
In momentum space, the two-point function of~$j_\mu$ is given by~\twopt\ with
\eqn\taufnphi{\eqalign{& \tau\left({p^2 \over m^2}\right) = {2 \over \pi } \left[\left(1+ {4 m^2 \over p^2} \right) {\rm arccot} \left({2 |m| \over |p|}\right) - {2 |m| \over |p|}\right]~,\cr
& \kappa = 0~.}}
The fact that~$\kappa = 0$ follows from parity. The function~$\tau(p^2/m^2)$ interpolates between~$\tau= 1$ in the UV and the empty theory with~$\tau = 0$ in the IR,
\eqn\tauvar{\tau\left({p^2 \over m^2}\right) = \cases{1 + \CO\left({|m| \over |p|}\right) &  \quad $p^2 \gg m^2$ \cr
{2|p| \over 3 \pi |m|} + \CO\left({|p|^3 \over |m|^3}\right) & \quad $p^2 \ll m^2$}}

Now consider a Dirac fermion~$\psi$ with real mass~$m$,
\eqn\psilag{{\scr L} = - i \b \psi \gamma^\mu \d_\mu \psi + i m \b \psi \psi~.}
The mass term explicitly breaks parity. The~$U(1)$ flavor symmetry that assigns charge~$+1$ to~$\psi$ gives rise to the current
\eqn\psicurr{j_\mu = - \b \psi \gamma_\mu \psi~,}
whose two-point function is given by~\twopt\ with
\eqn\taukapfnpsi{\eqalign{
& \tau\left({p^2 \over m^2}\right) = {2 \over \pi } \left[\left(1- {4 m^2 \over p^2} \right) {\rm arccot} \left({2 |m| \over |p|}\right) + {2 |m| \over |p|}\right]~,\cr
& \kappa\left({p^2 \over m^2}\right) = -{m \over
 |p|} \, {\rm arccot}\left({2 |m| \over |p|}\right)~.}}
Note that~$m\rightarrow -m$ under parity, so that~$\tau$ is invariant and~$\kappa \rightarrow -\kappa$. Again, the function~$\tau\left({p^2 / m^2}\right)$ interpolates between~$\tau = 1$ in the UV and~$\tau = 0$ in the IR,
\eqn\tauvarpsi{\tau\left({p^2 \over m^2}\right) = \cases{1 + \CO\left({m^2 \over p^2}\right) & \quad $ p^2 \gg m^2$ \cr
 {4 |p| \over 3 \pi|m| } + \CO\left({|p|^3 \over |m|^3}\right) & \quad $ p^2 \ll m^2$}}
The function~$\kappa\left({p^2 / m^2}\right)$ interpolates from~$\kappa = 0$ in the UV, where the theory is massless and parity invariant, to~$\kappa = -\half \sign (m)$ in the empty IR theory,
\eqn\kappavarpsi{\kappa\left({p^2 \over m^2}\right) =  \sign (m)  \cases{  -{\pi |m| \over 2 |p|} + \CO\left({m^2 \over p^2}\right)& \quad $ p^2 \gg m^2$\cr
 -\half  + \CO\left( \, {p^2 \over m^2}\right)& \quad $p^2 \ll m^2$}}

\ifig\taukapfree{Function~$\tau$ for the free scalar (blue,\ dotted) and functions~$\tau$ and~$|\kappa|$ for the free fermion (red,\ dashed and solid).}
{\epsfxsize=3.5in \epsfbox{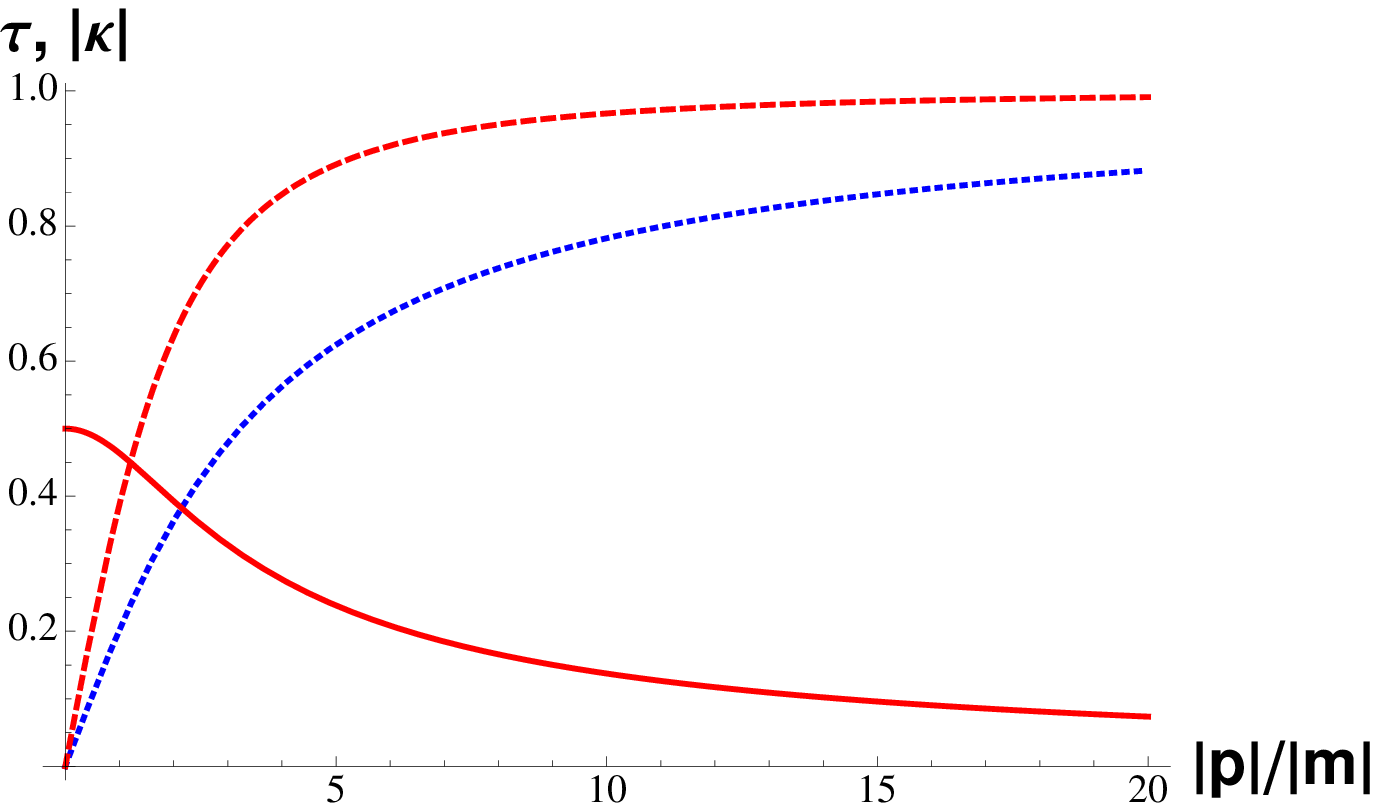}}

\noindent The function~$\tau\left({p^2 / m^2}\right)$ in~\taufnphi\ for a free scalar and the functions~$\tau\left({p^2 / m^2}\right)$ and~$\kappa\left({p^2 / m^2}\right)$ in~\taukapfnpsi\ for a free fermion are shown in~\taukapfree. At the scale~$p^2 \approx m^2$ these functions display a rapid crossover from the UV to the IR.

In theories with~$\CN=2$ supersymmetry, we can consider a single chiral superfield~$\Phi$ with real mass~$m$. This theory has a global~$U(1)$ flavor symmetry, and the associated conserved current~$j_\mu$, which resides in the real linear multiplet~$\CJ = \b \Phi e^{2 i m \theta \b \theta} \Phi$, is the sum of the currents in~\phicur\ and~\psicurr. Therefore, the function~$\tau\left({p^2 \over m^2}\right)$ is the sum of the corresponding functions in~\taufnphi\ and~\taukapfnpsi. Since~$\kappa\left({p^2 \over m^2}\right)$ only receives contributions from the fermion~$\psi$,  it is the same as in~\taukapfnpsi. From~\tauvar\ and~\tauvarpsi, we see that the total~$\tau\left({p^2 \over m^2}\right) \approx 2$ when~$p^2 \gg m^2$. In supersymmetric theories it is thus convenient to define~$\hat \tau = {\tau \over 2}$, so that that~$\hat \tau\left({p^2 \over m^2}\right) \approx 1$ when~$p^2 \gg m^2$ for a chiral superfield of charge~$+1$ and real mass~$m$.

\appendix{B}{Supergravity in Three Dimensions}

In this appendix we review some facts about three-dimensional~$\CN=2$ supergravity, focusing on the supergravity theory associated with the~$\CR$-multiplet. It closely resembles~$\CN=1$ new minimal supergravity in four dimensions~\SohniusTP. For a recent discussion, see~\refs{\KuzenkoXG,\KuzenkoRD}.

\subsec{Linearized Supergravity}

We can construct a linearized supergravity theory by coupling the~$\CR$-multiplet to the metric superfield~$\CH_\mu$,
\eqn\matcoup{ \delta {\scr L} = - 2 \int d^4 \theta \,  \CR_\mu \CH^\mu ~.}
The supergravity gauge transformations are embedded in a superfield~$L_\alpha$,
\eqn\tdmetgi{\eqalign{ \delta \CH_{\alpha\beta} = \half \left(D_\alpha \b L_\beta - \b D_\beta L_\alpha \right) + (\alpha \leftrightarrow\beta)~.}}
Demanding gauge invariance of~\matcoup\ leads to the following constraints:
\eqn\constraints{
 D^\alpha \b D^2 L_\alpha + \b D^\alpha D^2 \b L_\alpha = 0~.}
In Wess-Zumino gauge, the metric superfield takes the form\foot{Like the~$\CR$-multiplet in~\rmultcomp, the metric superfield contains factors of~$i$ that are absent in Lorentzian signature.}
\eqn\hcompbis{\CH_\mu = \half \left(\theta \gamma^\nu \b \theta\right) \left(h_{\mu\nu} - i B_{\mu\nu}\right) - {1 \over 2} \theta \b \theta C_\mu- {i \over 2} \thetasq \b \theta \b \psi_\mu +{i \over 2} \b \thetasq \theta \psi_\mu + \half \thetasq \b \thetasq \left(A_\mu-V_\mu\right)~.}
Here~$h_{\mu\nu}$ is the linearized metric, so that~$g_{\mu\nu} = \delta_{\mu\nu} + 2 h_{\mu\nu}$. The vectors~$C_\mu$ and~$A_\mu$ are Abelian gauge fields, and~$B_{\mu\nu}$ is a two-form gauge field. The gravitino~$\psi_\mu$ will not be important for us.  We will also need the following field strengths,
\eqn\vhdef{\eqalign{& V_\mu = - \ep_{\mu\nu\rho} \d^\nu C^\rho~, \qquad \d^\mu V_\mu = 0~\cr
& H = \half \ep_{\mu\nu\rho} \d^\mu B^{\nu
\rho}~.}}
We can now express the coupling~\matcoup\ in components,
\eqn\matcoupcomp{\delta {\scr L} = - T_{\mu\nu} h^{\mu\nu} + j^{(R)}_\mu \big(A^\mu - {3 \over 2} V^\mu\big) - i j_\mu^{(Z)} C^\mu + J^{(Z)} H+ \left({\rm fermions}\right)~.}
Since the gauge field~$A^\mu$ couples to the~$R$-current, we see that the gauge transformations~\tdmetgi\ include local~$R$-transformations. This supergravity theory is the three-dimensional analog of~$\CN=1$ new minimal supergravity in four dimensions~\SohniusTP.

It will be convenient to introduce an additional superfield,
\eqn\vh{\CV_{\CH} = {1 \over 4} \gamma^{\alpha\beta}_\mu [D_\alpha, \b D_\beta] \CH^\mu~,}
which transforms like an ordinary vector superfield under~\tdmetgi. Up to a gauge transformation, it takes the form
\eqn\vhcomp{\CV_{\CH} = \left(\theta\gamma^\mu\b\theta\right) \big(A_\mu - \half V_\mu\big) - i \theta\b\theta H + {1 \over 4} \thetasq \b \thetasq \left(\d^2 {h^\mu}_\mu - \d^\mu \d^\nu h_{\mu\nu}\right) +\left({\rm fermions}\right)~.}
The corresponding field strength~$\Sigma_{\CH} = {i \over 2} D \b D \CV_{\CH}$ is a gauge-invariant real linear superfield. The top component of~$\CV_{\CH}$ is proportional to the linearized Ricci scalar,
\eqn\linric{R = 2 \left(\d^2 {h^\mu}_\mu - \d^\mu \d^\nu h_{\mu\nu}\right) + \CO\left(h^2\right)~.}
With this definition, a~$d$-dimensional sphere of radius~$r$ has scalar curvature~$R = -{d(d-1) \over r^2}$.

In a superconformal theory, the~$\CR$-multiplet can be improved to a superconformal multiplet with~$\CJ^{(Z)} = 0$, as discussed in subsection~4.2. In this case the superfield~$L_\alpha$ is no longer constrained by~\constraints, and hence~$\CH_\mu$ enjoys more gauge freedom. In particular, this allows us to set~$H$ and~$A_\mu-\half V_\mu$ to zero. The combination~$A_\mu - {3 \over 2} V_\mu$ remains and transforms like an Abelian gauge field.

\subsec{Supergravity Chern-Simons Terms}

We will now derive the Chern-Simons terms~\lcs,~\ggcs, and~\frcs\ in linearized supergravity. We begin by considering terms bilinear in the gravity fields,
\eqn\gravterms{\delta {\scr L} = -2  \int d^4 \theta \, \CH^\mu\CW_\mu(\CH)~.}
Here~$\CW_\mu(\CH)$ is linear in~$\CH$. By dimensional analysis, it contains six supercovariant derivatives. Comparing to~\matcoup, we see that~$\CW_\mu(\CH)$ should be invariant under~\tdmetgi\ and satisfy the defining equation~\rmult\ of the~$\CR$-multiplet. It follows that the bottom component of~$\CW_\mu(\CH)$ is a conserved current.

There are two possible choices for~$\CW_\mu(\CH)$,
\eqn\wbas{\eqalign{&\CW_\mu^{(g)} = i \left(\delta_{\mu\nu} \d^2 - \d_\mu \d_\nu \right) \b D  D \CH^\nu + {1 \over 4} \gamma^{\alpha\beta}_\mu [D_\alpha, \b D_\beta] \Sigma_{\CH}~,\cr
& \CW_\mu^{(zz)} = {1 \over 8} \gamma_\mu^{\alpha\beta} [D_\alpha, \b D_\beta] \Sigma_{\CH}~.}}
The first choice~$\CW_\mu^{(g)}$ leads to the~$\CN=2$ completion of the gravitational Chern-Simons term~\lcs,
\eqn\lcsder{\eqalign{{\scr L}^{(g)} =  {i \over 4} \ep^{\mu\nu\rho} \Tr \big(\omega_\mu \d_\nu \omega_\rho + {2 \over 3} \omega_\mu \omega_\nu \omega_\rho\big) + i \ep^{\mu\nu\rho} \big(A_\mu - {3 \over 2} V_\mu\big) \d_\nu \big(A_\rho - {3 \over 2} V_\rho\big) + \left({\rm fermions}\right)~.}}
Here~$(\omega_\mu)_{\nu\rho} = \d_\nu h_{\rho \mu} - \d_\rho h_{\nu\mu} + \CO(h^2)$ is the spin connection. Note that we have included terms cubic in~$\omega_\mu$, even though they go beyond second order in linearized supergravity, because we would like our final answer to be properly covariant. Both terms in~\lcsder\ are conformally invariant and only the superconformal linear combination~$A_\mu - {3 \over 2}V_\mu$ appears. This is due to the fact that~\lcsder\ is actually invariant under the superconformal gauge freedom~\tdmetgi\ without the constraint~\constraints.

Upon substituting the second choice~$\CW^{(zz)}_\mu$, we can integrate by parts in~\gravterms,
\eqn\ggintbp{{\scr L}^{(zz)} = - \int d^4 \theta \, \CV_{\CH} \Sigma_{\CH}~,}
to obtain the~$Z$-$Z$ Chern-Simons term~\ggcs,
\eqn\ggder{{\scr L}^{(zz)} =i \ep^{\mu\nu\rho} \big(A_\mu - \half V_\mu \big) \d_\nu \big( A_\rho - \half V_\rho\big) + \half H R + \cdots + \left({\rm fermions}\right)~.}
Here the ellipsis denotes higher-order terms in the bosonic fields, which go beyond linearized supergravity. This term contains the Ricci scalar~$R$, as well as~$H$ and~$A_\mu - \half V_\mu$, and thus it is not conformally invariant.

It is now straightforward to obtain the flavor-gravity Chern-Simons term~\frcs\ by replacing~$\Sigma_{\CH} \rightarrow \Sigma$ in~\ggintbp. This amounts to shifting the~$\CR$-multiplet by an improvement term~$\delta \CR_\mu = {1 \over 8} \gamma_\mu^{\alpha\beta} [D_\alpha, \b D_\beta] \Sigma$. In components,
\eqn\frder{{\scr L}^{(fr)} = {i \over 2} \ep^{\mu\nu\rho} a_\mu\d_\nu \big(A_\rho - \half V_\rho\big) + {1 \over 8} \sigma R - \half D H + \cdots + \left({\rm fermions}\right)~.}
As above, the ellipsis denotes higher-order terms in the bosonic fields and the presence of~$R$,~$H$, and~$A_\mu - \half V_\mu$ shows that this term is also not conformally invariant.

\listrefs
\end